\documentclass[sigconf,9pt]{acmart}

\usepackage{algorithm}
\usepackage{algpseudocode}
\usepackage{xcolor} % 用于颜色
\usepackage{amsmath}
\usepackage[english]{babel}
\usepackage{blindtext}
\usepackage{verbatim}
\usepackage{circledsteps}
\usepackage[labelformat=simple]{subcaption}

\usepackage{graphicx}
\usepackage{enumitem}
\usepackage{booktabs} % 提供更专业的横线控制
\usepackage{multirow} % 处理合并单元格
\usepackage{tablefootnote}
\usepackage{booktabs}
\usepackage{makecell}

\usepackage{pifont}
\usepackage[normalem]{ulem} % 用于划掉词句
\newcommand{\nickname}{UBEP\xspace}

\newcommand{\affnum}[1]{\texorpdfstring{$^{\text{#1}}$}{}}
\newcommand{\affstar}[1]{\texorpdfstring{$^{\text{#1,}*}$}{}}
\newcommand{\affdag}[1]{\texorpdfstring{$^{\smash{\text{#1,\textdagger}}}$}{}}

\setcopyright{acmlicensed}
\acmYear{2026}\copyrightyear{2026}
\acmConference[SIGCOMM '26]{ACM SIGCOMM 2026 Conference}{August 17--21, 2026}{Denver, CO, USA}
\acmBooktitle{ACM SIGCOMM 2026 Conference (SIGCOMM '26), August 17--21, 2026, Denver, CO, USA}
\acmDOI{10.1145/3789240.3829183}
\acmISBN{979-8-4007-2467-1/26/08}

\acmPrice{}

\begin{document}
\begin{sloppypar}
\title[\nickname]{\nickname: Re-architecting Expert Parallelism Communication Library for Production Superpods}

%\titlenote{Produces the permission block, and copyright information}
%\subtitle{Extended Abstract}

\author{
  Yipeng Liu\affstar{1}, Chang Liu\affstar{1}, Si Shen\affstar{1}, Jiaqi Zheng\affdag{1}, Mingfan Li\affnum{2}, Yuyang Yang\affnum{1}, Guanhua Li\affnum{1}, Yuquan Zhang\affnum{1},
  Yimeng Xu\affnum{1}, Zhongzhe Hu\affdag{2}, Zhiyuan Huang\affnum{2}, Qihang Duan\affnum{2},
  Junsong Wang\affnum{2}, Wenkai Ling\affnum{2}, Baochuan Yang\affnum{2}, Xianzhi Yu\affnum{2}, Han Bao\affnum{2}, Yijie Chen\affnum{2}, Guihai Chen\affnum{1}
}

\affiliation{%
  % \vspace{0.1cm}
  \institution{\affnum{1}State Key Laboratory for Novel Software Technology, Nanjing University \quad \affnum{2}Huawei Technologies Co., Ltd.}
  \country{}
}

% \hypersetup{%
%   pdfauthor={Yipeng Liu, Chang Liu, Si Shen, Jiaqi Zheng, Mingfan Li, Yuyang Yang, Guanhua Li, Yuquan Zhang, Yimeng Xu, Zhongzhe Hu, Zhiyuan Huang, Qihang Duan, Junsong Wang, Wenkai Ling, Baochuan Yang, Xianzhi Yu, Han Bao, Yijie Chen, and Guihai Chen}
% }

% 脚注部分：同时包含共同一作和通信作者的说明
\thanks{$^{\ast}$Equal contribution.}
\thanks{$^{\dagger}$Corresponding author.}

\renewcommand{\authors}{Yipeng Liu, Chang Liu, Si Shen, Jiaqi Zheng, Mingfan Li, Yuyang Yang, Guanhua Li, Yuquan Zhang, Yimeng Xu, Zhongzhe Hu, Zhiyuan Huang, Qihang Duan, Junsong Wang, Wenkai Ling, Baochuan Yang, Xianzhi Yu, Han Bao, Yijie Chen, and Guihai Chen}

\renewcommand{\shortauthors}{Y. Liu, C. Liu, S. Shen, et al.} % X.et al.

\begin{abstract}
The deployment of Mixture-of-Experts (MoE) models on production high-bandwidth superpods, such as NVIDIA's NVL72/576 and Huawei's CloudMatrix384, introduces critical challenges beyond raw interconnect bandwidth. While these systems provide unified global address spaces and high-bandwidth fabrics, their full potential for sparse MoE communication is hindered by three fundamental bottlenecks: (1) Strict execution serialization imposed by coarse-grained Bulk Synchronous Parallel (BSP) orchestration of interdependent communication phases; (2) Prohibitive synchronization overhead that fails to scale alongside high interconnect bandwidth; and (3) Severe load imbalance resulting from distance-agnostic scheduling of irregular token traffic.
To eliminate these bottlenecks, we introduce \nickname (Unified-Bus Expert Parallelism), a production-ready communication library that rethinks MoE’s All-to-All primitives for modern superpod architectures. 
Through large-scale experiments, \nickname reduces All-to-All latency by up to 52.4\% and MoE inference Time Per Output Token (TPOT) by up to 11.1\%.

\end{abstract}




% \makeatletter
% \def\@authorfont{\LARGE}
% \def\@affiliationfont{\LARGE}
% \makeatother

\maketitle

\section{Introduction}

The rapid evolution of Large Language Models (LLMs) has established the Mixture-of-Experts (MoE) architecture as the standard for balancing massive parameter scales with inference efficiency~\cite{deepseekv3_2024}.
To support these communication-intensive workloads, datacenter infrastructure is shifting from traditional commodity clusters toward specialized superpod architectures, such as NVIDIA’s NVL72/576~\cite{NVL576} and Huawei’s CloudMatrix384 (CM384)~\cite{ServingLLM384}.
Unlike traditional clusters that rely on scale-out networks such as InfiniBand (IB)/RoCE with high tail-latency and explicit message passing, superpods integrate hundreds of accelerators via scale-up interconnects such as NVLink/Unified-Bus (UB), forming a unified, high-bandwidth, load/store-accessible domain.

Despite the unprecedented raw capabilities of superpods, efficiently harnessing these capabilities for the irregular and sparse communication patterns inherent to MoE models remains a critical system challenge.
Central to this challenge is the Expert Parallelism Communication Library (EPCL), the system layer responsible for orchestrating fine-grained token exchange. 
Unlike generic Collective Communication Libraries (CCLs) such as NCCL, EPCLs like DeepEP~\cite{deepep2025} implement the All-to-All primitive through narrow dispatch/combine APIs tailored to modern MoE systems.

As summarized in Table \ref{tab:framework_comparison}, existing EPCLs are not designed for modern multi-tier superpods. 
DeepEP~\cite{deepep2025} and UCCL-EP~\cite{ucclep2025} adopt the Bulk Synchronous Parallel (BSP) model for non‑superpod architectures with hybrid interconnects (e.g., IB and NVLink) and coarse‑grained kernel‑level scheduling, where long transmission time dominates and hides software overhead.
CANN EP~\cite{ServingLLM384} runs on a superpod fabric but retains BSP with intra-only pipelines and global barriers, leaving software overhead exposed on a low-latency fabric.
Hybrid-EP~\cite{HybridEP2025arXiv} adopts Asynchronous Parallel (ASP) model and supports single-tier superpods with finer warp-level scheduling, but it does not account for hierarchical communication constraints. 
In modern multi-tier superpods, drastically reduced link latency exposes the software overhead of BSP-style serialization and kernel scheduling, resulting in severe underutilization of the high-bandwidth fabric.

In this work, we identify three fundamental bottlenecks in re‑architecting EPCL for modern superpods:
(1) \noindent\textbf{BSP-style Serialization.} 
MoE communication inherently involves interdependent phases like routing and reordering; existing BSP implementations rely on strict global barriers for consistency --- a design that serializes execution in low‑latency superpods~\cite{ServingLLM384,deepep2025}. This previously overlooked ``stop-and-wait'' behavior prevents overlapping independent communication phases and leaves high‑bandwidth interconnects underutilized during synchronization.
(2) \noindent\textbf{Synchronization Tax.} 
We reveal that as link bandwidth scales in superpods, the relative cost of synchronization primitives, such as flags, barriers, and kernel launches becomes dominant, a phenomenon we term the ``synchronization tax''. Traditional control‑data decoupled mechanisms introduce non‑negligible overhead, which in ultra‑low‑latency environments limits scalability by dominating the end‑to‑end latency budget --- an issue not previously quantified in the context of modern EPCLs.
(3) \noindent\textbf{Topology-Agnostic Scheduling.} 
Although superpods provide a logically unified address space, we demonstrate that physical latency non‑uniformity across switching tiers remains a critical yet neglected factor. Current EPCLs treat the fabric as flat and schedule workloads based solely on token counts, ignoring heterogeneous access costs in multi‑tier fabrics. We show that this topology‑agnostic approach leads to severe stragglers and degraded tail latency --- a mismatch between logical abstraction and physical reality that has not been systematically addressed in prior EPCL designs.

To address these bottlenecks, we introduce \nickname, a production-ready EPCL designed specifically for modern superpods. Departing from the traditional BSP model, \nickname adopts a dependency‑driven execution model (\S\ref{sec:kernel_decomposition}) that decomposes the monolithic All‑to‑All primitive into fine‑grained tasks scheduled by data availability instead of global barriers. This enables aggressive overlap of metadata exchange, token dispatch, and reordering.
To mitigate the synchronization tax, we propose Data-as-Flag (\S\ref{sec:sync}), a novel mechanism that embeds synchronization signals directly into data payloads via atomic instructions. This allows implicit, near‑zero‑overhead coordination at the token level, effectively eliminating control‑plane overhead.
Finally, to overcome the limitations of topology‑agnostic scheduling, \nickname employs a hierarchical token-level scheduler (\S\ref{sec:scheduling})  that jointly optimizes token‑to‑core mapping by considering both load balance and physical fabric distance across switching tiers, thereby minimizing stragglers and tail latency.

We implemented \nickname on the Huawei CANN stack and evaluated it on up to 256 NPU dies allocated from a production CM384 superpod.
The evaluation results demonstrate that \nickname reduces All-to-All latency by 52.4\% compared to the baseline CANN EP, translating to an 11.1\% improvement in end-to-end Time Per Output Token (TPOT) for models at the scale of DeepSeek‑R1.

\noindent\textbf{Contributions.} Our primary contributions are as follows.
\begin{itemize}[leftmargin=*]
\item We characterize bottlenecks in BSP‑based EPCLs deployed on superpods, highlighting how synchronization overhead and serialization limit bandwidth utilization.
\item We design and implement \nickname, an EPCL natively for modern multi-tier superpods, featuring three core innovations: (1) token‑level kernel decomposition to maximize parallelism, (2) a hierarchical scheduler that mitigates tail latency from fabric heterogeneity, and (3) the Data‑as‑Flag mechanism to minimize synchronization overhead.
\item We evaluate \nickname on a production-scale CM384 superpod, demonstrating significant latency reduction and end‑to‑end performance gains for MoE inference.
\end{itemize}

\begin{table}[t]
  \centering
  \caption{Comparison of MoE communication libraries.}
  \label{tab:framework_comparison}
  
  % 核心修改：使用 resizebox 包裹 tabular
  % \linewidth 表示当前栏的宽度，! 表示高度自动保持比例
  \resizebox{\linewidth}{!}{
  \begin{tabular}{l c c c c c}
    \toprule
    \textbf{Framework} &
    \textbf{Net.} &
    \textbf{Target Arch.} &
    \textbf{Pipeline} &
    \textbf{Sync.} &
    \textbf{Sched.} \\
    \midrule

    DeepEP\cite{deepep2025} &
    IB+NV &
    Non-Superpod &
    Inter+Intra &
    BSP &
    Kernel \\
    
    UCCL-EP\cite{ucclep2025} &
    Hete\tablefootnote{It supports running on heterogeneous GPUs and NICs.} &
    Non-Superpod &
    Inter+Intra &
    BSP &
    Kernel
    \\

    Hybrid-EP\cite{HybridEP2025arXiv} &
    IB+NV &
    1-Tier-Superpod &
    Inter+Intra &
    \textbf{ASP} &
    \textbf{Warp} \\

    CANN EP\cite{ServingLLM384} &
    \textbf{UB} &
    2-Tier (Base) &
    \textbf{Intra-only}\tablefootnote{UB enables a large-scale superpod execution without inter-node communication.} &
    BSP &
    Kernel \\

    \midrule

    \textbf{UBEP (Ours)} &
    \textbf{UB} &
    \textbf{2-Tier (Opt)} &
    \textbf{Intra-only} &
    \textbf{ASP} &
    \textbf{Core} \\

    \bottomrule
    \vspace{-2.em}
  \end{tabular}
  } % resizebox 结束大括号
\end{table}

\noindent\textbf{Ethics.} This work does not raise any ethical issues.

\section{Background}
\subsection{Model Evolution}
\noindent\textbf{From Dense Model to Sparse Model.}
The MoE paradigm has emerged as a dominant strategy for scaling large models effectively. By activating only a small, input-specific subset of its total parameters (called \textit{experts}), MoE architectures enable a dramatic increase in model size without a proportional increase in computational cost, thereby improving overall capacity and performance. Consequently, many state-of-the-art models have adopted this sparse approach~\cite{deepseekv3_2024,qwen32025,glm452025}.
This architectural shift, however, introduces a fundamental change in communication patterns. Unlike the predictable, structured communication in dense model parallelism, MoE layers generate highly irregular and dynamic inter-GPU communication.
During inference or training, tokens are dynamically routed to experts via a learned gating function, with each GPU hosting only a subset of the total expert pool. This necessitates a two-step redistribution: first, token activations must be dispatched across the fabric to reach their assigned experts, and later, the processed outputs must be combined and returned to their original GPUs. 
The result is a sparse, runtime-dependent all-to-all exchange that forms the core communication bottleneck of MoE models~\cite{DeepSeekinsights2025}, with such communication consuming $\sim$47\% of the total execution time on average~\cite{comet2025, speculative-moe2025}.

\subsection{Architectural Evolution}
\begin{figure}[t]
  \centering
    \includegraphics[width=\linewidth]{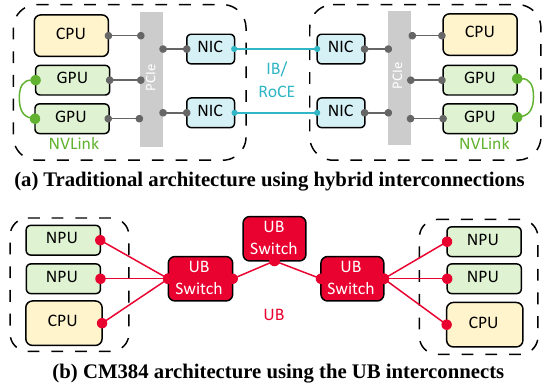}
  \vspace{-1.em}
  \caption{Comparison of traditional cluster using hybrid interconnects (a) with modern superpod utilizing the UB (b).}
  \label{fig:cm_384}
  \vspace{-0.5em}
\end{figure}

\noindent\textbf{Beyond Traditional Clusters: The Era of Modern Superpods.} 
Platforms like NVIDIA's NVL72~\cite{NVL576} and Huawei's CM384~\cite{ServingLLM384} illustrate the progression from traditional clusters with multi-node 8-accelerator systems to modern superpods featuring 72- or 384-accelerator integrated fabrics.
As illustrated in Figure \ref{fig:cm_384}(a), traditional clusters typically employ a hybrid interconnect strategy: 
NVLink for intra-node GPU-to-GPU communication, and IB/RoCE for inter-node networking. 
While this architecture supports cluster-scale communication through RDMA, its network bandwidth and topology are primarily optimized for data or pipeline parallelism (DP/PP), which generate relatively modest inter-node traffic~\cite{ServingLLM384}. 
In contrast, tensor parallelism (TP) and expert parallelism (EP) demand frequent, fine-grained, and low-latency communication, a requirement that is difficult to satisfy efficiently across traditional cluster nodes~\cite{ServingLLM384}. Consequently, many deployments are forced to confine TP/EP groups within a single compute node, constraining the scalability of the model. 
% \todo{above can be compact?}
Modern superpods, such as those realized by NVL72 and CM384, address this bottleneck by integrating hundreds of  accelerators into a single, coherent fabric. These systems exhibit three defining characteristics that are crucial for enabling efficient communication for MoE: (1) Modern superpods adopt advanced interconnect protocols, including NVLink~\cite{nvlink}, UB~\cite{UBspec}, Infinity Fabric~\cite{InfinityFabric2021}, UALink~\cite{UALink2025}, and SUE~\cite{BroadcomSUE2024}, which provide high bandwidth and low latency. For example, Huawei’s UB protocol in CM384 removes redundant network layers (Figure~\ref{fig:cm_384}(b)) and creates a direct connection between the GPU I/O Die and the on-chip Network-on-Chip (NoC), delivering bandwidth nearing 400~GB/s with latencies in the range of hundreds of nanoseconds. Furthermore, these fabrics employ scalable topologies (e.g., Mesh or CLOS) to aggregate multiple ports, ensuring non-blocking, high-bandwidth connectivity across the entire system~\cite{liao2025ub}.
(2) Unified Global Address Space (UGAS): superpod provides a unified memory address space with coherent load/store semantics, where all interconnected devices are mapped into a single, globally unique address domain. This architectural feature enables direct, universal memory access, which is essential for the software synchronization required by fine-grained parallel strategies. 
(3) Multi-Level, Pooled Resource Management: the superpod implements multi-level resource pooling, abstracting distributed compute, memory, and network resources into a cohesive logical pool. This allows for dynamic and flexible scheduling tailored to workload demands. 

\begin{figure}[t]
  \centering
  \includegraphics[width=\linewidth]{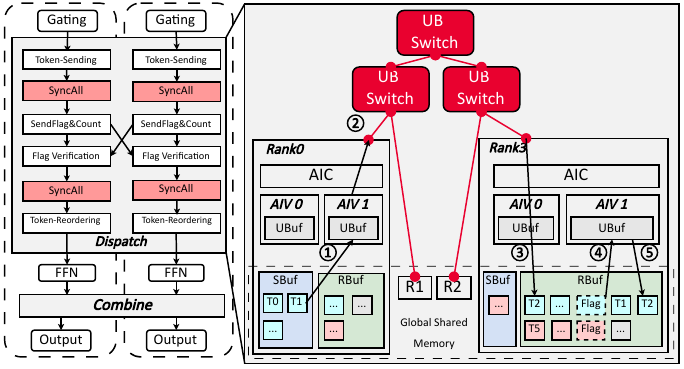}
  \vspace{-1.em}
  \caption{MoE Dispatch workflow. 
  Left: logical view of sequential phases with global barriers (SyncAll). Right: physical view on CM384.}
  \label{fig:moe}
  \vspace{-1.em}
\end{figure}

\subsection{Deploying MoE on CM384: Key Insights}

Modern superpods offer significant architectural advantages over traditional clusters; however, the direct deployment of MoE models onto these systems often exposes new challenges.

\noindent\textbf{Terminology.}
CM384 is built from a two-tier switching fabric: NPUs on the same baseboard connect through first-tier switches, while multiple baseboards connect through second-tier switches.
We classify memory accesses by whether they stay within the local NPU or traverse the switching fabric.
Since each NPU contains two compute dies, \textit{Intra-NPU} refers to cross-die HBM access within the local NPU.
We do not model same-die HBM access as a separate class, as its latency is negligible compared with accesses that traverse the switching fabric. \textit{One-Hop} and \textit{Two-Hop} denote NPU-to-NPU accesses that traverse one and two switch layers, respectively.
Orthogonal to this topology terminology, we use AIC and AIV to describe the execution hierarchy within each NPU.
Each NPU contains multiple AI Cores (AICs), and each AIC follows a decoupled 1-to-N design: one Cube Unit for matrix multiplication and multiple Vector Cores (AIVs) for parallel vector processing are orchestrated by a Scalar Unit that manages instruction dispatch and control flow.

\noindent\textbf{Deploying MoE on CM384.}
Typically, MoE layer involves two tightly coupled communication stages: (1) a dispatch stage, which routes tokens to remote GPUs hosting the target experts, and (2) a combine stage, which gathers expert outputs and restores the original token order.
Figure~\ref{fig:moe} details the typical execution flow of the dispatch operation on CM384.
Prior to execution, each rank allocates a global shared memory region partitioned by source rank and expert.
Leveraging the UGAS, the destination rank determines specific memory ranges based solely on per-expert token counts from each source rank.
The specific execution steps for the dispatch operation are as follows: 
(1) Initialization (Phase 1): Tokens are stored in the local Unified Buffer (UBuf) of each AIV.
(2) Transmission (Phase 2): Each AIV sends all locally owned tokens to their designated expert locations. A single token may be dispatched to multiple experts, while each expert receives tokens from multiple AIVs.
(3) Global Synchronization and Completion Signaling (Phase 3): A \texttt{SyncAll} operation ensures all token data has been written to global memory. AIVs responsible for signaling transmit per-expert flags and token counts to indicate processing readiness.
(4) Flag Verification (Phase 4): Verification is performed in parallel; each AIV polls a subset of flag entries in global memory, followed by another \texttt{SyncAll} to confirm global completion.
(5) Offset Calculation \& Reordering (Phase 5): Based on the finalized token counts, the system computes reordered token offsets and performs a contiguous reordering of tokens in global memory.
This intricate process highlights how the transition from dense to sparse modeling not only changes computational patterns but also demands synchronization-heavy communication protocols to manage the resulting irregular data movement efficiently.

\begin{figure}[t]
    \centering
    \begin{subfigure}[t]{0.19\textwidth}
        \centering
        \includegraphics[width=\linewidth]{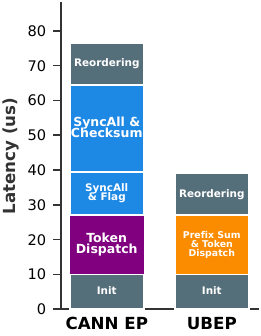}
        \caption{Latency breakdown of distinct phases across CANN EP, and \nickname.}
        \label{fig:perf_comparison}
    \end{subfigure}
    \hfill
    \begin{subfigure}[t]{0.27\textwidth}
        \centering
        \includegraphics[width=\linewidth]{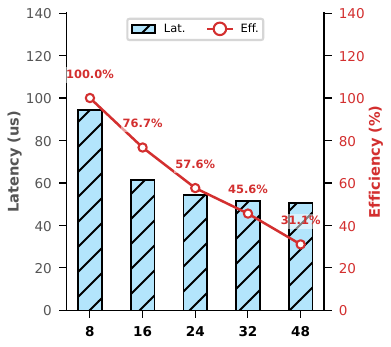}
        \caption{Scalability of the Token-Sending phase with varying AIV allocations.}
        \label{fig:scaling}
    \end{subfigure}
    
    \caption{Performance analysis of All-to-All communication.}
    \label{fig:perf_evaluation}
    \vspace{-1.em}
\end{figure}

\noindent\textbf{Our Key Insights.} 
In our effort to optimize communication for MoE models on modern superpods, we have re-architected the EPCL. Through this reconstruction, we derived three key insights that guided our architectural decisions and implementation.

\label{key_insight1}
\noindent\textbf{Insight 1: The BSP model's implicit synchronization incurs explicit barriers, creating a critical bottleneck for superpod. Unlocking their potential requires fine-grained task decomposition to maximize parallelism.}
The BSP model organizes MoE communication into a sequence of distinct phases, where each phase is guarded by a global synchronization barrier. While this design ensures correctness, it enforces strict ordering constraints and requires every AIV to perform the exact same type of work within a phase. This approach offers little flexibility in scheduling work across AIVs, resulting in a rigid execution path for the current dispatch. As shown in Figure~\ref{fig:moe}, phases such as \texttt{SetFlagAndCount} and \texttt{Token-Reordering} occur in a strict sequence, and a new phase can only begin once every AIV has reached the synchronization point. Since computation is assigned statically, faster AIVs often sit idle to wait for slower ones, leading to underutilized resources.

Figure~\ref{fig:perf_comparison} shows that this inefficiency becomes much more visible on modern superpods. In traditional clusters, where scale-out bandwidth was limited (50 GB/s for IB/RoCE v2), the extended time required for data transfer meant that synchronization overhead was simply less pressing compared to the transfer costs. 
On the CM384 architecture, however, the bandwidth increases by over an order of magnitude to 392 GB/s. As a result, token dispatch reaches bandwidth saturation very quickly, leaving phase-level synchronization as the dominant constraint.
We observe that only a subset of AIVs is sufficient to saturate the available bandwidth during dispatch, and adding more AIVs beyond this point provides little latency benefit.
As illustrated in Figure~\ref{fig:scaling}, scaling communication cores from 24 to 48 offers minimal benefit, because the extra cores spend most of their time waiting at barriers rather than doing useful work. This suggests an opportunity to re-evaluate the BSP-style execution. Since we do not need every AIV to participate uniformly in every phase, it is possible to break down these rigid boundaries and assign sub-tasks at a finer granularity. This would relax the strict dependencies and allow for more flexible scheduling.

\noindent\textbf{Insight 2: The explicit synchronization overhead on modern superpod architectures can dominate execution time. Leveraging hardware-guaranteed atomic operations enables implicit synchronization, replacing costly software barriers with memory-level consistency and forming the foundation for fine-grained, low-overhead parallel communication.}
While decomposing sub-task dependencies improves task parallelism, explicit synchronization primitives limit overall efficiency due to their overhead. On modern superpod architectures, this is exemplified by operations such as \texttt{SyncAll}, whose intrinsic cost can account for approximately 15\% of total execution time.
To address this, we leverage hardware-guaranteed atomic operations—specifically the 512B atomic Load/Store capability supported by the Ascend NPU within the superpod across scale-up and NoC domains. Crucially, the UB Fabric ensures atomicity of these writes and serializes subsequent reads, enabling \textit{implicit synchronization}. This approach replaces costly explicit software barriers with memory-level consistency, forming the foundation for truly fine-grained, low-overhead parallel communication primitives.

\begin{table}[t]
  \centering
  \caption{CM384 memory-access latency hierarchy under distinct distance categories. \textit{Intra-NPU} denote access HBM in local NPU; \textit{One-Hop} and \textit{Two-Hop} denote NPU-to-NPU accesses that traverse one and two switch layers, respectively.}
  \label{tab:memory_latency}
  \begin{tabular}{c c c}
    \toprule
    \textbf{Memory Access Level} & \textbf{Latency (ns)} & \textbf{Normalized} \\
    \midrule
    Intra-NPU & 218 & $1.0\times$ \\
    One-Hop & 929 & $4.3\times$ \\
    Two-Hop & 2500 & \textbf{$11.5\times$} \\
    \bottomrule
  \end{tabular}
  \vspace{-1em}
\end{table}

\noindent\textbf{Insight 3: In the hierarchical network of a modern superpod, load-balancing tokens without considering the significant latency gap between one-hop and two-hop memory accesses is counterproductive, resulting in stragglers that degrade overall MoE performance.}
Modern superpod architectures are rapidly evolving from scales of hundreds of accelerators (e.g., NVL72) to thousands (e.g., CM384, NVL576).
To support this massive scale, the interconnect fabric inevitably expands from a single-tier switch to a multi-tier switching architecture.
Consequently, while these superpods continue to provide a unified global address space, the underlying physical topology re-introduces distinct Non-Uniform Memory Access (NUMA) characteristics at the cluster level.
However, existing implementations remain oblivious to this shift: they are either tailored for single-tier architectures (e.g., NVL72)~\cite{HybridEP2025arXiv} or focus exclusively on balancing token counts~\cite{ServingLLM384}, failing to account for the heterogeneous access costs in multi-tier fabrics.
This mismatch leads to severe stragglers in \texttt{Token-Sending} phase because, in such hierarchical architectures, NPU-to-NPU memory access falls into three distinct distance categories with markedly different latencies:
(1) \textit{Intra-NPU} access to cross-die HBM within the local NPU (e.g., Rank 0 writing to HBM on another die of its local NPU);
(2) \textit{One-Hop} access across a single switch layer (e.g., Rank 0 to Rank 1); and
(3) \textit{Two-Hop} access traversing two switch layers.
As shown in Table~\ref{tab:memory_latency}, the latency gap is profound, with Two-Hop access reaching up to 11.5$\times$ that of local access, necessitating a scheduling approach that optimizes for both workload distribution and hierarchical memory access distances.

\section{\nickname}
\subsection{Overview}

\begin{figure}[t]
    \centering
    \includegraphics[width=\linewidth]{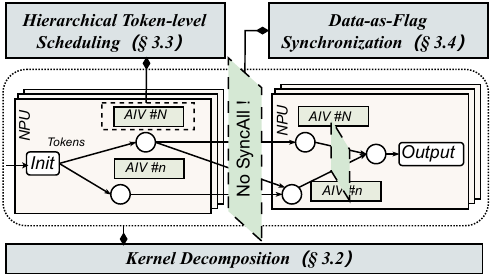}
    \caption{Overview of the \nickname\ architecture. }
    \label{fig:overview}
\end{figure}

\nickname (Unified-Bus Expert Parallelism) is a communication library for low-latency MoE inference.
We build \nickname on three key designs (Figure~\ref{fig:overview}):
First, we introduce \textbf{Kernel Decomposition} (\S\ref{sec:kernel_decomposition}) to break the sequential constraint of traditional BSP models by partitioning AIVs into distinct groups to execute independent tasks in parallel and replace global barriers with lightweight point-to-point synchronization.
Second, to handle the complex latency differences in superpods, we implement a \textbf{Hierarchical Token-level Scheduling} (\S\ref{sec:scheduling}). This component utilizes a hardware-accelerated mapper to generate optimal allocation schedules within 1~$\mu s$.
Finally, to eliminate expensive inter-NPU synchronization, we propose \textbf{Data-as-Flag} (\S\ref{sec:sync}).
This mechanism leverages the hardware's native 512-byte atomic load/store support. By embedding control flags directly within the data payload, \nickname achieves implicit consistency without separate control messages.
These three mechanisms are synergistic: kernel decomposition exposes the fine-grained tasks that the scheduler must balance across cores and that Data-as-Flag can synchronize without global barriers; without decomposition there would be no fine-grained work to overlap, without scheduling the workload would become unbalanced across the fabric, and without Data-as-Flag global barriers would reintroduce the synchronization tax.

In principle, the proposed techniques of \nickname's fine-grained parallelism can be adapted to any BSP-kernel, including both dispatch and combine routines. Since the challenges of global synchronization overhead and load imbalance are predominantly concentrated in the earlier dispatch phase, which fundamentally limits the overall scalability.  In this context, the following sections will take dispatch as an illustrative example,  without elaborating on trivial yet intricate details in combine.

\subsection{Kernel Decomposition}
\label{sec:kernel_decomposition}

\begin{figure*}[t]
  \centering
  \setlength{\abovecaptionskip}{0pt}
  \includegraphics[width=\linewidth]{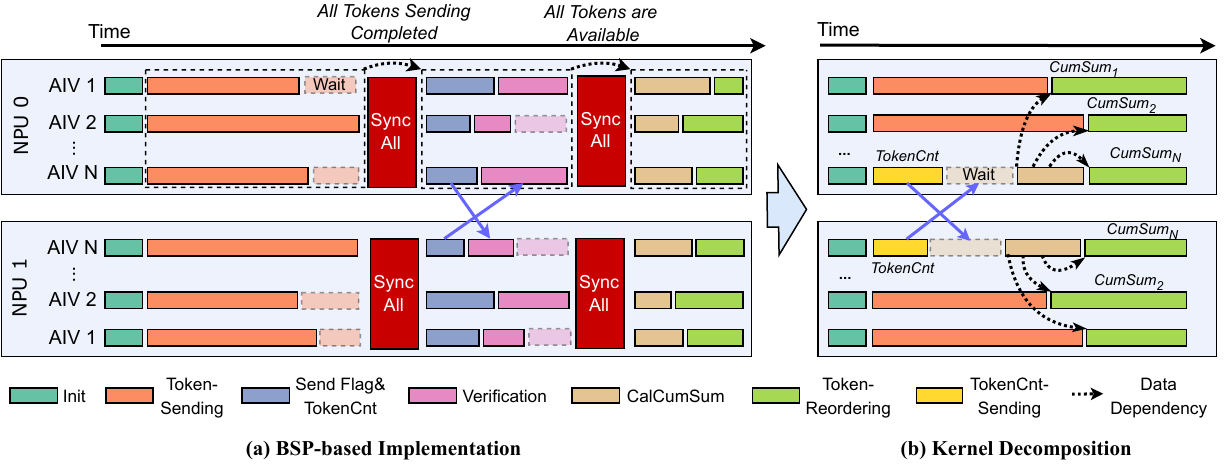}
  \vspace{-1.em}
  \caption{Comparison of BSP-based multi-phase execution and kernel decomposition for token dispatch.}
  \label{fig:decompose_kernel}
  \vspace{-0.5em}
\end{figure*}

Current MoE dispatch primitive organize communication and data reordering as BSP-style kernel executed. As illustrated in Figure~\ref{fig:decompose_kernel}(a), the traditional implementation simply divides the communication workflow into task phases executed in parallel across multiple AIVs. This simplifies programming and synchronization across AIVs by employing global barriers (\textit{i.e.,} \texttt{SyncAll}) to satisfy data dependencies: (1) All AIVs must wait for sending tokens (\textit{i.e.,} \texttt{Token-Sending}) before sending flags for tokens verification and token counts used for address calculation needed by reordering tokens (\textit{i.e.,} \texttt{Token-Reordering}); (2) All AIVs must verify all tokens before performing address calculation (\textit{i.e.,} \texttt{CalCumSum}). However, as each AIV follows a sequential execution order, tasks without dependencies cannot be performed in parallel, thereby restricting the potential for parallelism. Additionally, global barriers cause faster AIVs to wait in idle state during dispatch communication, which reduces effective utilization of NPU resources.

% 提出解决方案
Based on Insight 1 (\S\ref{key_insight1}), we observe that half of AIVs in each NPU are sufficient to saturate the transmission bandwidth. Therefore, we introduce kernel decomposition involving a redesign of the task decomposition for the entire communication workflow. In this approach, rather than requiring all AIVs to execute the same task sequence, we design a fine-grained partitioning mechanism that assigns independent tasks, such as token transmission and metadata process, to distinct AIVs for execution in parallel. Furthermore, we employ lightweight asynchronous signals to handle the synchronization across different AIVs.
Specifically, in Figure~\ref{fig:decompose_kernel}(b), each NPU decomposes \texttt{Token-Sending} and token counts sending (\textit{i.e.,} \texttt{TokenCnt-Sending}) tasks into two distinct AIV groups: most AIVs send tokens, while other AIVs calculate and send token counts.
Once all \texttt{TokenCnt} metadata become available, the latter AIVs compute prefix sums to determine HBM offsets for per-expert tokens used in \texttt{Token-Reordering}.
This decomposition allows us to overlap the processing latency of \texttt{TokenCnt} without compromising the bandwidth for token transmission.

Further, maximizing overlap and minimizing communication latency poses a constrained problem: given a fixed total number of AIVs, we must determine the optimal allocation ratio between these two groups. In general, the number of AIVs allocated to process token counts is proportional to the number of experts (calculating token counts for more destinations). Meanwhile, the number of AIVs assigned to token transmission is proportional to the product of the batch size and the Top-$k$ (handling larger communication volume). Detailed cost modeling and the derivation of this optimal allocation appear in Appendix~\ref{sec:lantency_model}. We also evaluate and verify the model’s impact in Appendix~\ref{sec:hot_figure}.

A data dependency exists within the entire workflow: \texttt{Token-Reordering} relies on the addresses calculated by \texttt{CalCumSum}. Rather than employing global synchronization via \texttt{SyncAll}, we implement point-to-point synchronization based on asynchronous data signals. Specifically, the AIVs tasked with \texttt{CalCumSum} write the calculated addresses into global shared memory and other AIVs poll for corresponding memory offset and execute \texttt{Token-Reordering} immediately upon detecting the update.

\subsection{Hierarchical Token-level Scheduling}
\label{sec:scheduling}

As highlighted in Insight 3, the structural disparity across multiple tiers induces latency asymmetry. In practical scheduling, this asymmetry manifests primarily between one-hop access and two-hop access traversing multiple switch layers; consequently, we omit the negligible impact of intra-NPU access. Furthermore, employing naive load balancing strategies in token dispatch often exacerbates the straggler problem (as detailed in \S \ref{sec:ablation_study}). Consequently, we must rethink token scheduling by establishing a unified model that considers the impact of load balancing and hierarchy.

\noindent\textbf{Formulation and Hardness Analysis.} 
Consider a specific NPU within a superpod that outputs \( m \) tokens, denoted by the set 
$\mathcal{T} = \{1, \dots, m\}$, 
which must be dispatched to \( n \) AIVs, represented by 
$\mathcal{C} = \{1, \dots, n\}$.
The goal is to determine an optimal assignment of tokens to AIVs such that the overall communication and processing latency is minimized, subject to satisfy load-balancing and assignment constraints.
Denote by \( \ell_i \) the transmission latency of the token \( i \) and by \( RTT_i \) the round-trip time between the token \( i \) and its destination (relevant when network delays are considered). Binary variables \( \{ x_{ij}\ |\ i \in \mathcal{T}, j \in \mathcal{C} \} \) indicate the assignment between tokens and AIVs; \( x_{ij} = 1 \) if and only if token \( i \) is assigned to AIV \( j \); and the set of tokens assigned to AIV \( j \) is defined as $\mathcal{M}_j = \{ i \mid x_{ij} = 1 \}$.
The objective is to minimize the maximum completion time across all AIVs, which consists of the cumulative transmission latency of assigned tokens and the worst-case network delay within each AIV’s token set.
\begin{subequations}
\vspace{-1em}
\label{eq:opt_model} 
\begin{align}
    \text{minimize} \quad & \max_{j \in \{1,\dots,n\}} \left( \sum_{i \in \mathcal{M}_j} \ell_i + \max_{i \in \mathcal{M}_j} \{ \text{RTT}_i \} \right) \tag{\ref{eq:opt_model}} \\
    \text{subject to} \quad & \sum_{j=1}^n x_{ij} = 1, \quad \forall i \in \{1, \dots, m\} \label{eq:c1} \\
    & \sum_{i=1}^m x_{ij} \le \left\lceil \frac{m}{n} \right\rceil, \quad \forall j \in \{1, \dots, n\} \label{eq:c2} \\
    & x_{ij} \in \{0, 1\}, \quad \forall i, j \label{eq:c3}
\end{align}
\end{subequations}

The constraint \eqref{eq:c1} characterizes that each token is assigned to exactly one AIV.  
The binary decision variable $x_{ij}$ indicates whether the token \( i \) is assigned to AIV \( j \). 
Load balancing is enforced by the constraint \eqref{eq:c2} so that no AIV receives more than a fair share of tokens.
This formulation captures both the communication overhead (via \( \ell_i \)) and the network heterogeneity (via \( RTT_i \)), while ensuring that the workload distribution remains balanced across AIVs—a key requirement for scalable and low-latency token dispatch in MoE-based architectures.
% 【NP-hard问题，证明见附录】
We formally prove that the program \eqref{eq:opt_model}  is NP-hard and the detailed proof is provided in Appendix~\ref{sec:hardness_proof}.

\noindent\textbf{Latency Homogenization.}
Directly solving this NP-hard optimization problem to optimality is prohibitively expensive in the context of our sub-microsecond latency budget. Even state-of-the-art integer programming solvers would require much more time to compute a solution, which would itself become the dominant bottleneck, utterly negating the performance gains sought from an optimized schedule.
We propose that the objective can be heuristically solved through latency homogenization:
making the latency composition of each AIV similar, thereby avoiding bottlenecks caused by some AIVs processing too many high latency tokens.
To approximate the NP-hard objective, our heuristic algorithm decomposes the problem based on the two additive terms in the program~\eqref{eq:opt_model}.

First, we consider the transmission time.
We note two key observations:
(1) in LLMs, tokens generally have uniform sizes across the model, which are much smaller than the bandwidth. Therefore, the transmission time per token (\( \ell_i \)) can be approximately treated as the same;
(2) in modern high-bandwidth superpods, the transmission time for a single token (typically at the KB level) is much shorter than the network propagation latency.
Based on these observations, we can propose that $\sum_{i \in \mathcal{M}_j} \ell_i \propto \sum_{i=1}^m x_{ij}$.
Therefore, this aspect can be satisfied together with constraint~\eqref{eq:c2}.
What's more, the second observation justifies why our subsequent optimization can focus primarily on the network propagation term in the objective function, as it becomes the dominant source of latency variance.

Second, we consider the RTT. 
We categorize tokens by hop count and aim to equalize the hop-count distribution across AIVs.
Let $v_j$ be the hop distribution vectors of the AIV $j$ and let $\Sigma$ be the their covariance matrix. Minimizing the variance of each hop-count component across AIVs is equivalent to minimizing $tr(\Sigma)$.
This reduces the heuristic to the simple per-hop capacity bound $v_{j,h} \leqslant \lceil V_h / n \rceil$, where $V_h$ is the total number of $h$-hop tokens.

\noindent\textbf{Hardware-Accelerated Mapper.} 
The theoretical formulation of the token scheduling problem provides a foundation for optimizing token assignments. 
Practical deployment, however, requires a mechanism that operates efficiently under hardware constraints. 
In this work, we propose a hardware-accelerated algorithm design that ensures load balancing over all tokens and maintains balance across categories.
The design comprises two tightly coupled components:
% The hardware-accelerated mapper comprises two components.
(1) Expert Remapping via Logical Matrix Transposition: We construct a virtual matrix where rows correspond to AIVs and columns to experts. By reading this matrix in column-major order, we generate a remapped expert sequence for each AIV. Using efficient vectorized matrix operations, we change the original token sequence that AIVs were responsible for sending, thereby ensuring that each AIV accesses a balanced mix of one-hop and two-hop experts, homogenizing the expected communication latency.  
(2) Token-level Load Partition: The global token sequence is partitioned into contiguous slices of equal size, and each AIV is assigned one slice. This guarantees that the maximum load difference between any two AIVs is at most one token. Since the remapped expert sequence produces non-uniform per-expert token counts, this is a prefix-sum search: given a target token index $T_{target}$, find the expert whose cumulative token range covers it. Each AIV performs this lookup via quaternary search (Algorithm~\ref{alg:quaternary_search}), which maintains a search window $[L, R]$ over the expert space and evaluates three equidistant pivots per iteration. The cumulative token counts at all three pivots are computed in parallel via a single SIMD \texttt{VectorCount} instruction. The algorithm then narrows the window to the quadrant containing $T_{target}$.Upon convergence, the left boundary $L$ gives the expert ID, and a final VectorCount yields the local offset within that expert's buffer. The search converges in 4--5 steps for typical expert counts and stays within 1~$\mu$s.

\begin{algorithm}[t]
\caption{Quaternary Search}
\label{alg:quaternary_search}
\begin{algorithmic}[1]
    \algtext*{EndWhile}
    \algtext*{EndIf}
    \renewcommand{\algorithmicrequire}{\textbf{Input:}}
    \renewcommand{\algorithmicensure}{\textbf{Output:}}
    
    \Require Reordered Expert List $\mathbf{E}$ (tensor), Target Global Token Index $T_{target}$, Total Number of Experts $N_{exp}$
    \Ensure Target Expert ID $Exp_{idx}$, Local Token Offset $Offset$

    \State \textcolor{gray}{$\triangleright$ Initialization}
    \State $[L, R] \gets [0, N_{exp}]; \quad C_L \gets 0; \quad Gap \gets \lceil (R - L) / 4 \rceil$

    \While{$Gap \ge 1$}
        \State \textcolor{gray}{$\triangleright$ Step 1: Define Pivots }
        \State $P_i \gets L + i \cdot Gap \quad \forall i \in \{1, 2, 3\}$

        \State \textcolor{gray}{$\triangleright$ Step 2: SIMD Parallel Counting }
        \State $C_i \gets \Call{VectorCount}{\mathbf{E} < P_i} \quad \forall i \in \{1, 2, 3\}$

        \State \textcolor{gray}{$\triangleright$ Step 3: Narrow down (Find target quadrant $q$)}
        \State \textbf{Define} $P_0 \gets L, \ P_4 \gets R, \ C_0 \gets C_L$ 
        \State Find smallest $q \in \{1, 2, 3\}$ such that $T_{target} < C_q$
        \State \textbf{if} no such $q$ exists \textbf{then} $q \gets 4$
        \State \textbf{Update:} $L \gets P_{q-1}; \quad R \gets P_q; \quad C_L \gets C_{q-1}$

        \State $Gap \gets \lceil (R - L) / 4 \rceil$
        \If{$R \le L$ \textbf{or} $Gap = 0$} 
            \State \textbf{break} 
        \EndIf
    \EndWhile

    \State \textcolor{gray}{$\triangleright$ Finalization}
    \State $Exp_{idx} \gets L$
    \State $C_{final} \gets \Call{VectorCount}{\mathbf{E} < Exp_{idx}}$
    \State $Offset \gets T_{target} - C_{final}$
    \State \Return $Exp_{idx}, Offset$
\end{algorithmic}
\end{algorithm}

\noindent\textbf{Generalization Beyond Two Tiers.}
While we evaluate CM384's two-tier topology, the scheduler readily extends to an arbitrary number of hop classes $H$ by applying the same per-hop capacity constraint, $\lceil V_h / n \rceil$. Provided the topology exhibits distinguishable latency between different hops, the scheduler effectively balances tokens across AIVs. On fabrics like CM384, where per-hop latency variance is large, the benefits of the scheduling strategy are most pronounced. 

\begin{figure*}[t]
  \centering
  \includegraphics[width=\linewidth]{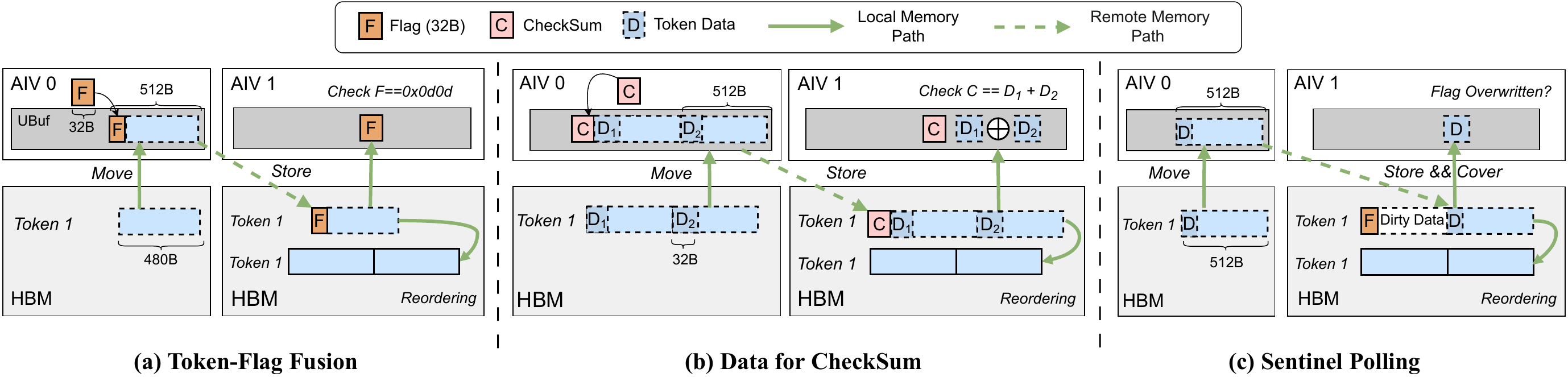}
  \vspace{-1.5em}
  \caption{Comparison of three Data-as-Flag–based synchronization mechanisms across NPUs.}
  \label{fig:data_as_flag_v1}
  \vspace{-0.5em}
\end{figure*}

\subsection{Data-as-Flag Synchronization}
\label{sec:sync}

Unlike strictly ordered systems, data transfer with memory semantic within the modern superpod necessitate explicit memory barriers to enforce visibility ordering between payload and flags, causing hardware pipeline stalls. To alleviate this, we propose Data-as-Flag, a lightweight synchronization scheme that ensures correctness through verification at a smaller scale by leveraging the 512 bytes atomic memory access capability of the CM384.

% 设计1
As illustrated in Figure~\ref{fig:data_as_flag_v1}(a), 
Token-Flag Fusion (TFF) encapsulates the payload and the synchronization flag within a single 512 bytes DataBlock, where the 32 bytes leading serves as the flag field and utilizing only the remaining 480 bytes for payload. 
Memory semantic enables direct load/store from the local memory of a source NPU to the HBM of a destination NPU. This mechanism necessitates the movement of token data from the HBM to the local memory, where a leading flag is naturally embedded to construct a 512 bytes DataBlock. Subsequently, each DataBlock is written atomically into the HBM of the receiver.
Verification of the flag alone is sufficient to ensure entire data transmission has been completed.
This approach eliminates the barrier on the sender side, thereby achieving barrier-free communication. However, embedding the flag within the DataBlock introduces an additional transmission overhead, which leads to a reduction in effective bandwidth.

% 设计2
To further reduce the overhead introduced by embedded flags, 
Data-for-Checksum (DC) completely eliminate explicit flags and instead leverage the token data itself for verification.
Specifically, as illustrated in Figure~\ref{fig:data_as_flag_v1}(b), the leading 32 bytes of each DataBlock is utilized as a verification marker.
Upon completing the token transmission, the sender calculates an accumulated checksum of these data markers and transmits it as a separate packet.
The receiver then uses this checksum to verify data integrity and confirm the completion of transmission.
However, while this approach improves bandwidth utilization, it necessitates waiting for the checksum of an entire batch, which hinders fine-grained, token-level pipelining.

% 设计3
Sentinel Polling (SP) further mitigates the overhead of flag generation and verification by shifting both operations entirely to the receiver, as illustrated in Figure~\ref{fig:data_as_flag_v1}(c).
Before communication, the receiver initializes the receive buffer with a specific initial value, while the sender transmits raw payload without any pre-processing. 
The receiver then determines data arrival by verifying whether the designated bytes (\textit{e.g.,} 32~B) within DataBlocks differ from the initial value. 
Although this approach eliminates the need for flag transmission and theoretically achieves 100\% effective bandwidth utilization, it necessitates active resetting of the receive buffer after data consumption.
More critically, a synchronization deadlock occurs if the transmitted data is identical to the initial value. While increasing the number of comparison bytes can mitigate the collision probability, it escalates the computational overhead, and this probability theoretically never reaches zero.

\noindent\textbf{Correctness and safety assumptions.}
Data-as-Flag relies on 512B atomicity (enforced by UB Fabric): A 512B DataBlock is written atomically and observed by remote NPUs as a single transaction.
Under this assumption, each Data-as-Flag variant admits a simple happens-before argument.
In TFF, the sender writes $(flag, payload)$ in one atomic 512B store; the receiver polls the flag and, once it changes, reads the same 512B block. Atomicity guarantees that the flag and payload are observed together, so $Write_{flag} \to Read_{payload}$.
In DC, the sender writes raw payload blocks and, after the batch completes, writes a checksum; the receiver waits for the checksum and then reads the payload. The checksum write acts as a per-batch barrier: $Write_{payload} \to Write_{checksum} \to Read_{payload}$.
In SP, the receiver pre-initializes the buffer with a sentinel value and the sender's atomic write of non-sentinel data overwrites it. Because the write is atomic, the receiver never observes a partially written DataBlock, so $Write_{data} \to Read_{data}$ unless the data itself equals the sentinel.

To eliminate SP's theoretical deadlock when payload matches the sentinel, we reserve a sentinel encoding that cannot be produced by normal computation (e.g., an unused bit pattern in BF16/FP16) and verify at model initialization that the MoE weights and activations never emit it.
With a 32B (256 bits) sentinel, the probability of accidental collision under a uniform value distribution is $2^{-256}$; in practice the bound is far lower because valid activations occupy a narrow subset of the value space.

\section{Implementation}
\label{sec:implementation}

We implemented \nickname\ based on the CANN software stack~\cite{huaweiCANN2025},
developing an optimized EPCL in approximately 10K lines of Ascend C. The implementation builds on UB primitives to leverage the available high-bandwidth, low-latency interconnect.

\noindent\textbf{SIMD Vectorized Operation.} 
We leverage the vector instruction set of Ascend NPU throughout \nickname's communication kernels to avoid scalar per-token loops. For the hierarchical mapper, \texttt{VectorCount} performs parallel prefix-sum lookups in a single instruction (Algorithm~\ref{alg:quaternary_search}). For the \texttt{Token-Sending} path, we use \texttt{CompareScalar} for masking and validation, and \texttt{FindFirstValue} to locate token indices.
This structure lets the kernel handle small batches efficiently while keeping the control flow straightforward.

\noindent\textbf{Global Load/Store for Point-to-Point Synchronization.} To support kernel decomposition, we replace BSP-based barriers with fine-grained point-to-point synchronization (\S\ref{sec:kernel_decomposition}). On conventional clusters, this style of synchronization is often limited by coarse kernel scheduling and by DMA overhead for small messages.
CM384 provides global addressing that supports cross-NPU notifications via instruction-level load/store. We use this capability both for synchronization and for data movement. During initialization, we pre-partition the receive address space, so senders can write tokens directly into the destination's global memory. This removes local aggregation and reduces shuffling on the critical path~\cite{fusco2025}.

\section{Evaluation}
\label{sec:evaluation}
In this section, we first examine \nickname's operator-level performance gains by focusing on core communication improvements~(\S\ref{sec:perf_comparison}).
We then demonstrate AIV-level pipelining enabled by \nickname's kernel decomposition~(\S\ref{sec:latency_breakdown}). Next, we analyze the hierarchical token-level scheduling and the synchronization mechanism~(\S\ref{sec:ablation_study}) through ablation studies. Finally, we demonstrate \nickname's end-to-end performance improvements across different models and verify the gains under various MoE model configurations~(\S\ref{sec:e2e_perf}).

\subsection{Setup}
\label{sec:setup}

% CM384
\noindent\textbf{Testbed.} We evaluate \nickname on a production-scale Huawei CM384 superpod. Our deployment consists of 16 Ascend servers allocated from this superpod, interconnected via a high-bandwidth, low-latency unified fabric, totaling 256 NPU Dies.

% 模型参数
\begin{table}[t]
    \centering
    \caption{\textbf{MoE Model configurations. R/S experts means routed and shared experts.}}
    \label{tab:model_config}
    \resizebox{\linewidth}{!}{ 
        \begin{tabular}{l|cc|cccc|cc}
            \toprule
            \textbf{Model Name} & \textbf{Total} & \textbf{Active} & \textbf{Layers} & \textbf{Hidden} & \textbf{Top-$k$} & \textbf{\makecell{Experts\\(R/S)}} & \textbf{Precision}\\
            \midrule
            \textbf{Qwen3-30B} & 30.5B & 3.3B & 48 & 2048 & 8 & 128/0 & BF16 \\ 
            \textbf{GLM-4.7} & 358B & 33.6B & 92 & 5120 & 8 & 160/1 & BF16 \\
            \textbf{DeepSeek-R1} & 671B & 37B & 61 & 7168 & 8 & 256/1 & W8A8 \\ 
            \textbf{DeepSeek-V3.2} & 685B & 37B & 61 & 7168 & 8 & 256/1 & W8A8 \\
            \bottomrule
        \end{tabular}
    }
    \vspace{-1em}
\end{table}

%选用的模型和参数
\noindent\textbf{Models and workload.} To demonstrate the versatility and robustness of \nickname across a wide spectrum of model architectures, we evaluate our system using a diverse suite of four state-of-the-art LLMs, ranging from 30B to 685B parameters. Detailed specifications for all evaluated models are summarized in Table \ref{tab:model_config}. We deploy these models on the CM384 cluster, adjusting the parallelism strategies (TP, DP, EP) according to each model's scale to ensure optimal hardware utilization. 

% V1(CANN EP) & V2(UBEP w/o mapping)
\noindent\textbf{Baseline.} We evaluate \nickname against the following baselines:
\begin{itemize}[leftmargin=*]
\item CANN EP~\cite{ServingLLM384}: a DeepEP-like engineering adaptation to CM384, representing the conventional BSP paradigm used in existing MoE frameworks.
\item \nickname w/o mapping: An ablation variant that retains basic load balancing but removes the Tier-2 network-aware scheduling, used to verify the benefits of our topology-aware design.
\item Synchronization Variants: To isolate the impact of synchronization strategies, we compare the baseline Stop-and-Wait (SW) mechanism (used in CANN EP) against three variants of our Data-as-Flag design: Token-Flag Fusion (TFF), Data Checksum (DC), and Sentinel Polling (SP).
% Stop-and-Wait
\end{itemize}

% operator latency & end-to-end latency(TPOT)
\noindent\textbf{Metrics.} We evaluate \nickname\ with two key metrics: operator latency and end-to-end latency. We first conduct a fine-grained dispatch operator latency breakdown to analyze sub-process execution time and measure total operator latency to compare with baselines. Then we use Time Per Output Token (TPOT), a critical metric for user experience in LLM serving, to show how our optimizations lead to lower token delivery latency at scale.

% fig1_2_comm_stacked_row.pdf
\begin{figure*}[t]
  \centering
  \includegraphics[width=\linewidth]{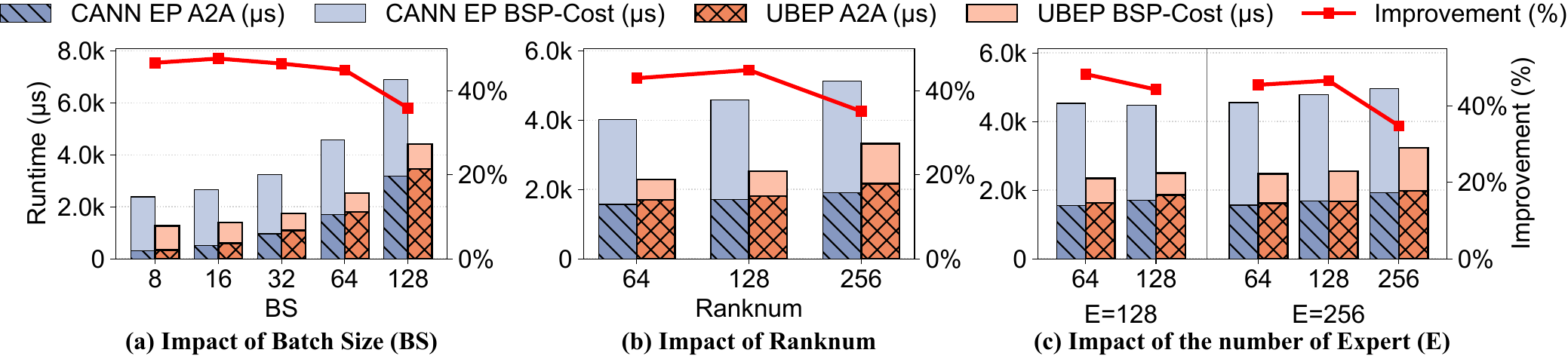}
  \vspace{-1.5em}
  \caption{Latency composition of the MoE dispatch operator across representative execution configurations.}
  \label{fig:perf}
  \vspace{-0em}
\end{figure*}

\subsection{Performance Evaluation}
\label{sec:perf_comparison}
\noindent\textbf{EP Performance Comparison.} 
We first evaluate the communication latency and effective bandwidth of \nickname in large-scale clusters.
Table~\ref{tab:ep_comparison} details the results.
\nickname improves performance by implementing kernel decomposition to increase parallelism, using Data-as-Flag mechanism to remove global synchronization barriers, and applying fine-grained AIV-level scheduling. Compared to CANN EP on identical hardware and topology, \nickname improves bandwidth by 35.3\%--40.8\%, isolating the algorithmic gains of our design.

% DeepEP & CANN EP & \nickname
\begin{table}[t]
    \centering
    \caption{Performance of different EPCLs. DeepEP is included as a protocol-capability reference on H800.}
    \label{tab:ep_comparison}
    \small % 保持字号适中
        \begin{tabular*}{\columnwidth}{@{\extracolsep{\fill}}ccccccc@{}}
        \toprule
        \multirow{4}{*}{\textbf{EP}} & \multicolumn{2}{c}{\textbf{DeepEP}\cite{deepep2025}} & \multicolumn{2}{c}{\textbf{CANN EP}\cite{ServingLLM384}} & \multicolumn{2}{c}{\textbf{\nickname}} \\ 
         & \multicolumn{2}{c}{(on H800)} & \multicolumn{2}{c}{(on CM384)} & \multicolumn{2}{c}{(on CM384)} \\
        \cmidrule(lr){2-3} \cmidrule(lr){4-5} \cmidrule(lr){6-7}
         & \textbf{Latency} & \textbf{BW} & \textbf{Latency} & \textbf{BW} & \textbf{Latency} & \textbf{BW} \\
         & ($\mu$s) & (GB/s) & ($\mu$s) & (GB/s) & ($\mu$s) & (GB/s) \\ \midrule
        % 8   & 77 & 98 & ? & ? & ? & ? \\
        16  & 118 & 63 & 103 & 71 & 73 & 100 \\
        32  & 155 & 48 & 120 & 61 & 86 & 85 \\
        64  & 173 & 43 & 139 & 53 & 101 & 73 \\
        128 & 192 & 39 & 144 & 51 & 106 & 69 \\
        256 & 194 & 39 & 151 & 48 & 112 & 66 \\ \bottomrule
        \end{tabular*}
    \vspace{0em}
\end{table}

\begin{figure*}[t]
  \centering
  \includegraphics[width=\linewidth]{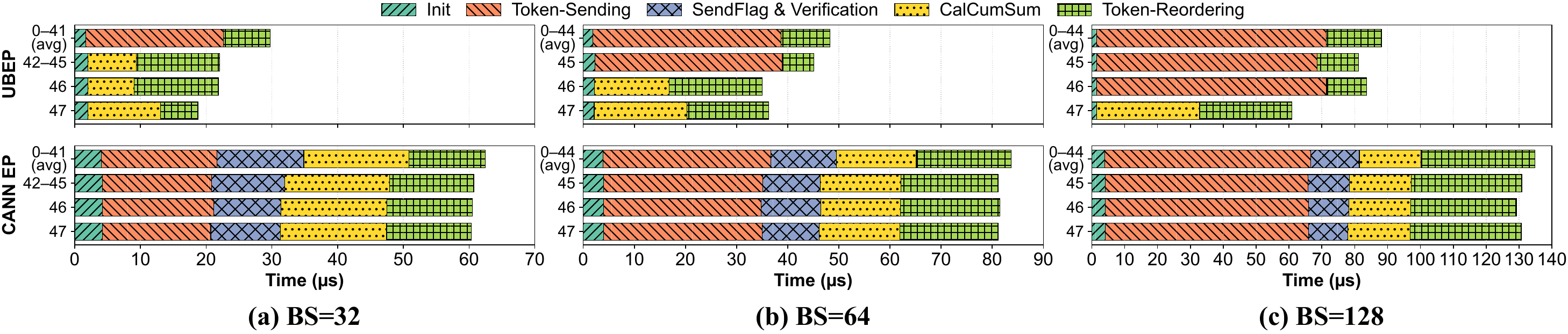}
  \vspace{-1.5em}
  \caption{Task-wise latency breakdown of the dispatch operator across varying batch sizes.}
  \label{fig:breakdown}
  \vspace{-1.em}
\end{figure*}

% Compare to CANN EP
\noindent\textbf{Sensitivity to Workload Parameters.}
To analyze the performance characteristics of \nickname in more detail, we evaluate it under different batch sizes, cluster scales, and expert counts. Unless otherwise stated, experiments use 128 ranks, a batch size of 64, and 1 expert per NPU. A more comprehensive analysis across different parameters is deferred to Appendix \ref{sec:appendix_latency_eval} and ~\ref{sec:additional_experimental_parameters} due to space constraints.

Figure~\ref{fig:perf}(a) shows that \nickname demonstrates consistent improvements across BS ranging from 8 to 64. By replacing the baseline mechanism with fine-grained, token-level pipelining, \nickname achieves an average gain of approximately 46.4\%. 
However, at $BS=128$, the gain decreases to 35.9\%. This decline is primarily due to the increased communication volume associated with larger batch sizes. Although we minimize the kernel resources for the \texttt{TokenCnt} calculation to a single core, the computation time remains shorter than the communication latency. Consequently, the computation phase cannot fully mask the expanding communication overhead, which limits the overall speedup.

% rank
% 1、rank增加会提升控制处理的复杂性，CANN EP不如UBEP；
% 2、rank增加会提升通信量（13%），此时计算核数量增多就会导致通信核减少，造成性能提升下降；
Figure~\ref{fig:perf}(b) illustrates how \nickname scales with the number of ranks. In the baseline, overheads for routing and global checksums grow with the rank number, accumulating latency due to repeated global barriers. \nickname mitigates this by pipelining the preprocessing steps, maintaining an average advantage of 43.9\% as the cluster scales.
As the rank number increases to 256, the improvement drops to 35.1\%. Two factors contribute to this. First, the communication volume increases by 13.2\%. Second, to handle the more complex routing and global checksums at this scale, we must allocate 12 compute cores for these tasks. This reduces the number of cores available for communication. The reduced parallelism in communication, combined with the larger data volume, prolongs the total transmission time.

% expert
Figure~\ref{fig:perf}(c) highlights the benefits of \nickname under complex routing logic. Increasing the number of experts exacerbates load imbalance. In the Baseline, straggler effects cause synchronization overheads to consume around 70\% of the total runtime. \nickname restricts this overhead to approximately around 30\%, yielding speedups of 38.6--41.9\%. As the number of ranks increases, the performance improvement is consistent with our previous analysis.
It is worth noting that \nickname exhibits slightly longer raw communication latency compared to the Baseline. This is because we divert specific AIV resources to perform \texttt{CalCumSum} through kernel decomposition, leaving fewer cores for data transmission. However, this decision proves highly beneficial: the minor increase in transmission time is far outweighed by the drastic reduction in synchronization costs, resulting in a net decrease in end-to-end latency.

\subsection{Latency Breakdown}
\label{sec:latency_breakdown}
To evaluate the effect of Kernel Decomposition, we conducted the tests with a fixed EP scale of 64 NPUs, using a configuration of one expert per NPU. Figure~\ref{fig:breakdown} illustrates the detailed latency breakdown of the dispatch operator for three specific batch sizes. In CANN EP, execution proceeds through five stages: \texttt{Init}, \texttt{Token-Sending}, \texttt{SendFlag \& Verification}, \texttt{CalCumSum} and \texttt{Token-Reordering}. Global barriers need to be inserted between the two phases where data dependencies exist, causing stragglers to increase the latency of the entire operator.
\nickname removes these global barriers by combining Kernel Decomposition with Data-as-Flag synchronization, reducing the idle tome of AIVs. Furthermore, fine-grained kernel decomposition enables parallel execution of \texttt{Token-Sending} and \texttt{CalCumSum}. Since the number of AIVs assigned to token transmission scales in proportion to the sending volume, the AIVs available for \texttt{CalCumSum} decrease as the batch size increases. The experiment results indicate that \nickname overlap the processing latency of \texttt{CalCumSum} with the communication latency of \texttt{Token-Sending} across different batch sizes, resulting in a performance improvement from 34.7\% to 52.4\% over CANN EP. We also evaluated optimal decomposition strategies under different EP scales, and the detailed experimental results are provided in the Appendix~\ref{sec:hot_figure}.

\subsection{Ablation Study}
\label{sec:ablation_study}
We conduct an ablation study to examine the contribution of individual design components in \nickname.

\begin{figure}[t]
  \centering
  \includegraphics[width=\linewidth]{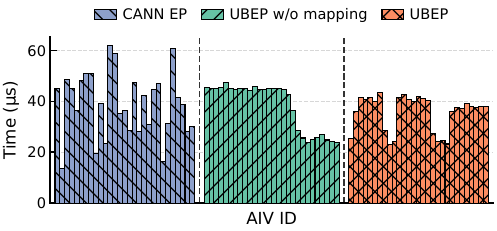}
  \vspace{-1.5em}
  \caption{AIV-Level token dispatch latency under different scheduling schemes.}
  \label{fig:ablation_token_scheduling}
  % \vspace{-1.em}
\end{figure}

\noindent\textbf{Impact of Token Scheduling.}
This study isolates the impact of token scheduling on dispatch latency, focusing on expert load imbalance and cross–Tier-2 communication. We compare \nickname with CANN EP and \nickname w/o mapping on a 64-NPU system with 256 experts and Top-$k$=8, organized as four 16-NPU nodes connected via a Tier-2 switch with eight hot experts. Figure~\ref{fig:ablation_token_scheduling} shows the token transmission latency of 28 representative AIVs selected from the 48 AIVs on a single node with 16 NPUs. CANN EP exhibits a highly skewed latency distribution due to hotspot experts that concentrate token sending on a small subset of AIVs, leading to a high maximum AIV latency of 62.2~$\mu s$. \nickname w/o mapping redistributes token sending load but remains topology-agnostic, causing most transmissions to traverse the Tier-2 switch. As the cross–Tier-2 latency is higher than local HBM access (Table~\ref{tab:memory_latency}), the AIV-level latency remains uneven, with a maximum latency of 48.1~$\mu s$. In contrast, \nickname jointly balances token sending load and aligns token scheduling with topology, effectively eliminating expert hotspots and latency imbalance caused by cross-switch traffic, achieving near-uniform token transmission latency across AIVs and reducing the maximum latency to 43.5~$\mu s$.

\begin{figure}[t]
  \centering
  \includegraphics[width=\linewidth]{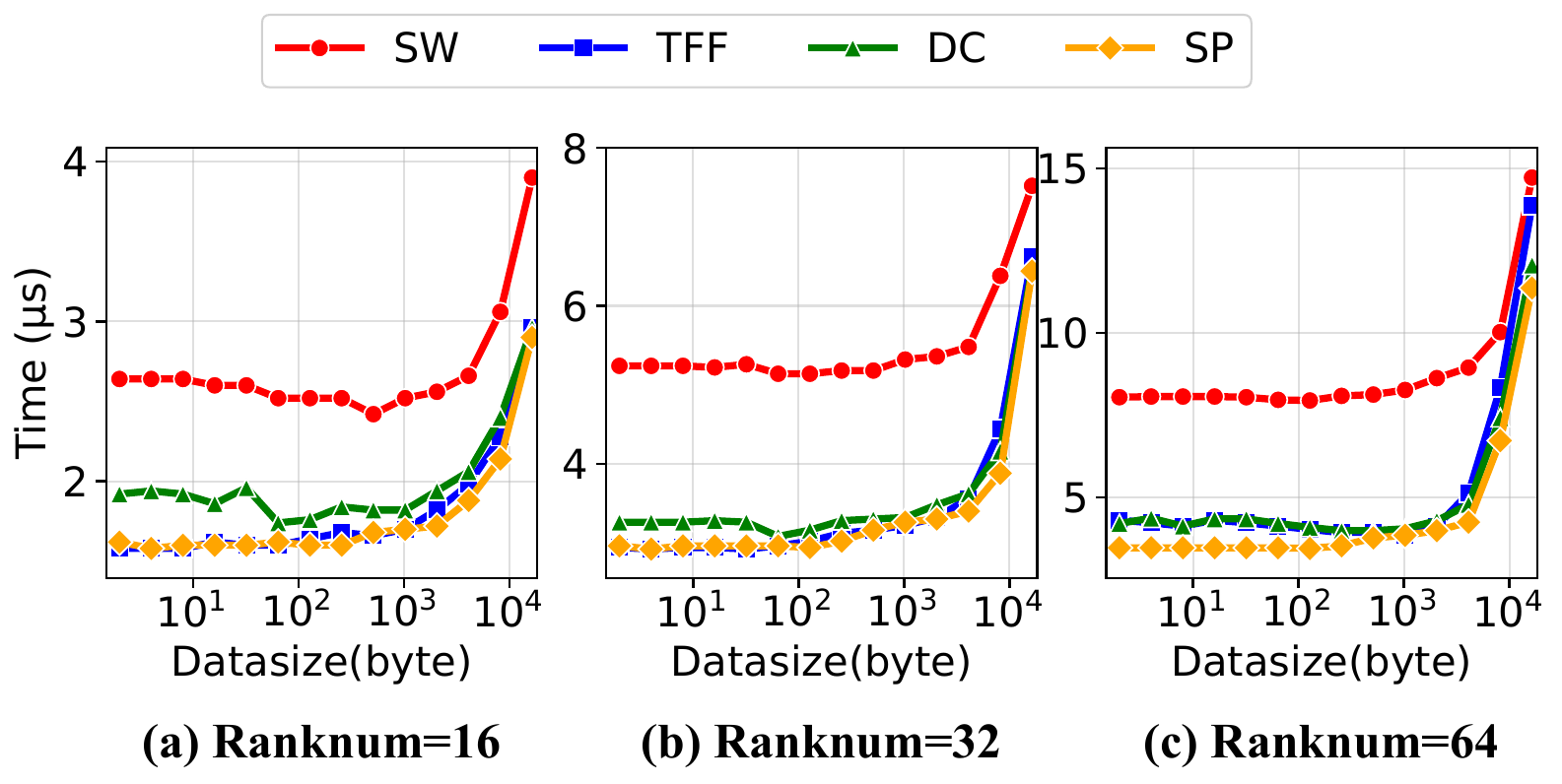}
  \caption{Dispatch latency of Data-as-Flag variants versus the CANN EP across varying synchronization granularity and ranks.}
  \label{fig:data_dataasflag}
  \vspace{-1.em}
\end{figure}

% Data-As-Flag消融实验
\noindent\textbf{Impact of Data-as-Flag.}
% We evaluate the TFF, DC, and SP variants against the CANN EP baseline across 
We evaluate three Data-as-Flag variants (TFF, DC, and SP) against CANN EP (SW) under different synchronization granularities and NPU scales.
Figure~\ref{fig:data_dataasflag} compares CANN EP with these variants. Across different granularities and rank numbers, all Data-as-Flag methods reduce latency up to 31.0\%–57.1\% relative to CANN EP, mainly by removing the stop-and-wait execution pattern and reducing cross-AIV synchronization overhead.
On the other hand, The benefit decreases as synchronization granularity increases. At a typical dispatch granularity of around 10~KB, the gain stays at about 1~$\mu s$. This occurs because as the synchronization granularity increases, the communication time accounts for a larger proportion of the total time, which gradually reduces the relative gain obtained from optimizing synchronization.
Furthermore, as the number of ranks increases, the synchronization overhead grows. Consequently, the absolute performance gain of TFF, DC, and SP reaches 3.7--4.6~$\mu s$ when the data volume is small.
Among the three variants, DC shows higher latency under small granularity or fewer NPUs because it waits for token batches, which weakens token-level pipelining. SP better balances bandwidth utilization and pipelining, and delivers lower latency across most settings, although it may encounter a rare deadlock if the sentinel value matches actual data.

% sec2：e2e performance
\subsection{End-to-End Inference Performance}
\label{sec:e2e_perf}

\begin{figure}[t]
  \centering
  \includegraphics[width=\linewidth]{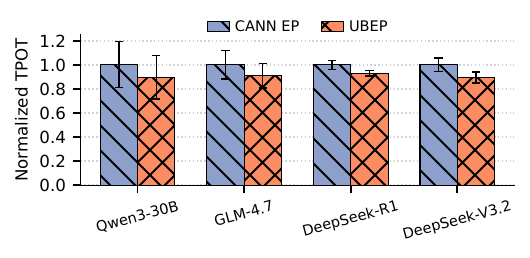}
  \vspace{-2.em}
  \caption{Normalized TPOT performance of CANN EP and \nickname}
  \label{fig:e2e_tpot}
  \vspace{-1.em}
\end{figure}

In this section, based on the vLLM-Ascend v0.14.0~\cite{vllmascend} (a hardware plugin for vLLM~\cite{vllm} on Ascend NPU), we evaluate the end-to-end inference performance of \nickname on  Mixture-of-Experts (MoE) large language models across varying scales, including Qwen3, GLM and DeepSeek. 
For Qwen3-30B model, we deploy it on 16 ranks.
GLM-4.7 are deployed on 128 and 160 ranks respectively, adopting one-expert-per-rank mapping.
DeepSeek-R1 and DeepSeek-V3.2 are evaluated on 128 ranks with a configuration of two experts per rank.

Figure~\ref{fig:e2e_tpot} reports the normalized end-to-end TPOT distribution across MoE models, with each per-token latency sample normalized to the corresponding CANN EP baseline.
\nickname reduces P99 end-to-end latency by up to 11.1\%, shifting the distribution downward, though the gain is smaller than the operator-level speedup because MoE All-to-All communication is only one component of the decoding step. Profiling further shows that MoE communication accounts for roughly 50\% of per-token latency (consistent with prior work~\cite{comet2025}) but only about 20\% of actual hardware execution time, suggesting substantial dependency stalls and runtime overhead. 
The remaining time is spent on FFN computation, attention, framework scheduling, kernel launch, and memory movement. 
These findings motivate future work to extend the same fine-grained dependency-driven machinery to other collective communication operations and computation.
Because model configurations lead to different operator-level characteristics, the normalized gains vary across models. 
Each box is collected from at least 100 decoding steps.

\section{Discussion}

\noindent\textbf{Portability and Hardware Assumptions.}
\label{sec:portability}
To help readers map \nickname onto other superpod fabrics, we explicitly separate the ideas that are fundamentally portable from those that rely on CM384-specific features.
The core ideas do not depend on the Ascend Instruction Set Architecture (ISA):
(1)~token-level kernel decomposition to expose fine-grained dependencies and overlap metadata preparation with data movement;
(2)~replacing global barriers with point-to-point data signals in globally-addressable memory;
(3)~topology-aware token scheduling that homogenizes the per-core hop-distance mix.
These ideas apply whenever the fabric provides a unified global address space with remote load/store and enough fine-grained concurrency units to exploit the exposed parallelism.

Three assumptions are currently tuned to CM384, summarized in Table~\ref{tab:portability}:
(1)~Data-as-Flag uses the 512B atomic write granularity provided by the UB fabric. On fabrics with smaller atomic units, \nickname can use smaller DataBlocks at lower payload efficiency; without atomic writes, it falls back to DC or explicit fences/acks;
(2)~\nickname uses AIV-level concurrency for fine-grained decomposition and overlap.
On GPU-like architectures, this role maps to warp- or thread-block-level specialization, including NVIDIA-style persistent kernels.
Unlike fused persistent kernels that often rely on rigid resource allocation and software-managed completion polling, \nickname uses receiver-driven data signals and instruction-level atomics to interleave communication and computation at sub-microsecond granularity;
(3)~the hierarchical scheduler assumes a multi-tier fabric with uniform bandwidth but hop-dependent latency. On flatter fabrics, it reduces to token-level load balancing with weaker straggler mitigation.

\begin{table*}[t]
  \centering
  \caption{Portability checklist for \nickname components.}
  \label{tab:portability}
  \small
  \setlength{\tabcolsep}{4pt}
  \renewcommand{\arraystretch}{1.08}
  \begin{tabular}{@{}m{0.2\linewidth} m{0.25\linewidth} m{0.5\linewidth}@{}}
    \toprule
    \textbf{Capability} &
    \textbf{Used by} &
    \textbf{Degradation if absent} \\
    \midrule

    AIV/warp-level concurrency
    & Kernel decomposition
      (\S\ref{sec:kernel_decomposition}, \S\ref{sec:scheduling})
    & Coarser scheduling; less overlap \\

    \addlinespace[1pt]

    Uniform-bandwidth fabric with hop-dependent latency
    & Hierarchical scheduler
      (\S\ref{sec:scheduling})
    & Falls back to load balancing; stragglers remain \\

    \addlinespace[1pt]

    512B atomic write
    & Data-as-Flag
      (\S\ref{sec:sync})
    & Smaller atomic blocks reduce efficiency; no atomicity requires DC or fences/acks \\

    \bottomrule
  \end{tabular}
\end{table*}

\noindent\textbf{Attention-FFN Disaggregation (AFD).} 
To address the resource dichotomy between memory-intensive Attention layers and compute-heavy FFN experts, recent architectures ~\cite{MegaScale-Infer2025, step3} propose AFD to physically decouple these components onto specialized clusters. This paradigm fundamentally alters the communication topology from symmetric All-to-All to a bipartite Many-to-Many (M2N) pattern, where nodes assume distinct sender or receiver roles. Crucially, our design philosophy remains invariant under this shift. Within the superpod context, \nickname's fine-grained, dependency-driven orchestration is topology-agnostic; it effectively overlaps the latency of M2N data transfers by decoupling synchronization from payload movement, regardless of the underlying traffic asymmetry.

\noindent\textbf{Next-Generation Fabric Semantics.}
We demonstrated the efficiency of AIV-level parallelism in \nickname on the CM384 superpod. As inference kernels enter the microsecond regime, breaking the global synchronous barriers of BSP becomes critical. 
While \nickname leverages UGAS to optimize bandwidth and hierarchical access, software overheads like Data-as-Flag~(\S3.4) still limit fine-grained scaling.
Offloading these control signals to dedicated hardware units represents a more promising mechanism—notable examples include the Scatter-Gather Engine within the SparseCore of TPUs (e.g., TPUv7~\cite{gcp2025tpu7x}) or the CC-Core in AWS Trainium~\cite{aws2025trainium2}.
Meanwhile, enforcing receiver-side ordering capabilities, like \textit{Write-with-Notify} or \textit{Remote-Sync-Write} semantics, can immediately trigger downstream computation pipelines. Representing the key evolutionary direction for UB, these capabilities would reduce programming complexity and 0.5 RTT overhead, unlocking greater potential for fine-grained computation-communication overlap.

\section{Related Work}

\noindent\textbf{Communication Optimization.} Efficient All-to-All communication is pivotal for MoE scalability~\cite{DeepSpeed_MoE2022, GLaM_PMLR2022, MegaScale-MoE2025, LSH_MoE2024NeurIPS}. 
A large body of work has optimized this primitive through specialized EPCLs or general-purpose communication runtimes. EPCLs such as DeepEP~\cite{deepep2025} and UCCL-EP~\cite{ucclep2025} target traditional IB/NVLink clusters with coarse-grained, BSP-style kernels.                         
Hybrid-EP~\cite{HybridEP2025arXiv} adopts asynchronous parallelism for single-tier superpods but does not address hierarchical topologies. 
FUSCO~\cite{fusco2025} reduces overhead by fusing layout handling, yet it does not target the superpod. 
More general runtimes such as NVSHMEM~\cite{NVSHMEM} and MSCCL++~\cite{MSCCLPP} support flexible data movement but rely on explicit ordering between data movement and notification (e.g., software fences, work-queue entries, or separate flag updates). 
These designs---whether MoE-specific or generic---target traditional scale-up/scale-out pipelines or flat single-tier superpods; they do not exploit the unified memory semantics and hierarchical topology of modern multi-tier superpods, where traffic patterns and latency non-uniformity render existing adaptation techniques ineffective~\cite{NVSwitch_HotChips2022, ScaleupSurvey2025}.

\noindent\textbf{Computation-Communication Overlapping.} Hiding latency via concurrency is a standard optimization in MoE systems~\cite{GPipe_NeurIPS2019,PipeDream_SOSP2019,Alpa2022_USENIX,Lina_USENIX2023,comet2025,fastermoe2022,pipemoe2023,flashmoe2025,schemoe2024,zheng2026uniepunifiedexpertparallelmoe,tutel_Mlsys2023,Lancet_MLSYS2024,hiermoe2025}. 
One line of work schedules decomposed All-to-All operations alongside expert computation, as in FasterMoE~\cite{fastermoe2022}, PipeMoE~\cite{pipemoe2023}, and Comet~\cite{comet2025}. Another line relies on persistent or fused kernels to blur the boundary between communication and computation. FlashDMoE~\cite{flashmoe2025} uses persistent kernels for device-initiated asynchronous communication and fine-grained pipelining, while UniEP~\cite{zheng2026uniepunifiedexpertparallelmoe} integrates dispatch, grouped GEMM, and combine within MoE megakernels. 
System-level frameworks such as Tutel~\cite{tutel_Mlsys2023}, Lancet~\cite{Lancet_MLSYS2024}, and HierMoE~\cite{hiermoe2025} optimize scheduling and communication, but target conventional GPU clusters with NVLink+IB interconnects rather than unified-memory superpods. 
These overlap strategies, whether based on data slicing, persistent kernels, or megakernels, require resolving complex data dependencies and allocating runtime resources precisely. The resulting contention and synchronization overheads can negate the intended speedups~\cite{ByteScheduler_SOSP2019, FuseCompute2024SC, CCFuser_ppopp2025}. 
Unlike prior work that overlaps EP communication with expert computation, \nickname hides latency within the EP communication primitive itself by decomposing it into data-dependency-aware sub-tasks, such as overlapping address calculation with token transmission.

\noindent\textbf{Load Balance.} Prior works primarily address straggler effects at the \textit{algorithm-level}~\cite{GShard_2021gshard,SwitchTransformers2022,SparselyGatedMoE_ICLR2017,tutel_Mlsys2023,NetMoE_ICLR2025,speculative-moe2025,eplb,lplb,gracemoe,dancemoe,ec2022}. Approaches include dynamic routing optimization~\cite{eplb, lplb, NetMoE_ICLR2025}, expert replication~\cite{gracemoe, dancemoe}, and inverted routing paradigms~\cite{ec2022}, which ensure balance at the cost of model fidelity or logic modification. For instance, EPLB~\cite{eplb} and LPLB~\cite{lplb} mitigate skew by dynamically altering token assignments based on global statistics, effectively reshaping the traffic pattern at the application layer. In contrast, \nickname targets \textit{system-level} orchestration. We accept the irregular traffic patterns generated by these upper-layer strategies and optimize the underlying dataflow and synchronization mechanisms to maximize effective bandwidth utilization on high-performance superpods without altering model semantics.

\section{Conclusion}

In this paper, we presented \nickname, a production-ready communication library re-architected for the era of superpods. 
\nickname dismantles the rigid BSP execution model by decomposing the monolithic All-to-All primitive into dependency-driven tasks. To eliminate explicit software barriers, we propose the Data-as-Flag synchronization protocol that leverages hardware-native atomic semantics. Furthermore, our hierarchical token-level scheduling mechanism neutralizes the load imbalance among AIVs and the straggler effects caused by hierarchical network latencies. 
Deployed on the Huawei's CM384 superpod, \nickname demonstrates significant performance gains, reducing All-to-All latency by up to 52.4\% and improving end-to-end MoE inference TPOT by up to 11.1\%. These results validate that to fully unleash the potential of next-generation AI superpods, communication software should evolve from coarse-grained orchestration to fine-grained parallelism.

\section*{Acknowledgments}

We sincerely thank all the anonymous reviewers and our shepherd for their helpful comments on drafts of this paper. The work is partly supported by the Fundamental and Interdisciplinary Disciplines Breakthrough Plan of the Ministry of Education of China (JYB2025XDXM901) and  the NSF of China (62422207).
\bibliographystyle{ACM-Reference-Format}
\bibliography{reference}
\clearpage
\appendix

\section*{Appendices}

Appendices are supporting material that has not been peer-reviewed.

\section{Analytical Modeling of Dispatch Latency}
\label{sec:lantency_model}

% 我们将dispatch分为Count计算和token分发两部分，并希望找到执行时间最短的分核方式。
We model the MoE dispatch process as two parallel groups executing on partitioned hardware resources: \textbf{token count processing} and \textbf{token dispatching}. To minimize the total latency, we aim to find the optimal resource partitioning that balances the execution time of these two groups.

%代价建模
\textbf{Cost Models.} Let $n$ denote the total number of available AIVs. We partition these resources into $n_{count}$ AIVs for processing token counts and $n_{token}$ cores for dispatching token data, such that $n_{token} = n - n_{count}$. We define the time cost functions for both groups as follows:
\begin{itemize}
    \item \textbf{Token Count Processing ($T_{count}$):} The latency is proportional to the number of experts ($N_{exp}$) and inversely proportional to the allocated compute resources ($n_{count}$).
    \begin{equation}
        T_{count} = G\left(\frac{N_{exp}}{n_{count}}\right) + b_{count}
    \end{equation}
    
    \item \textbf{Token Dispatching ($T_{token}$):} The latency is proportional to the total token load, defined by the product of the Top-$k$ value ($K_{top}$) and the batch size ($B_{sz}$), and inversely proportional to the allocated communication resources ($n - n_{count}$).
    \begin{equation}
        T_{token} = F\left(\frac{K_{top} \times B_{sz}}{n - n_{count}}\right) + b_{token}
    \end{equation}
\end{itemize}

Here, $F(\cdot)$ and $G(\cdot)$ represent the linear scaling functions of communication and computation, respectively. $b_{token}$ and $b_{count}$ denote the constant hardware overheads (e.g., kernel launch latency).

\textbf{Optimization Objective.} Since the two AIV groups execute in parallel, the total dispatch latency $T_{dispatch}$ is determined by the slower group (the bottleneck). Our objective is to solve the Min-Max problem:

\begin{equation}
    \min_{n_{count}} \Big( \max(T_{count}, T_{token}) \Big)
\end{equation}

According to the pipeline parallelism principle, the optimal latency is achieved when the two groups are perfectly overlapped, \textit{i.e.,} $T_{count} = T_{token}$. Substituting the cost models, we get:

\begin{equation}
    \label{eq:balance}
    G\left(\frac{N_{exp}}{n_{count}}\right) + b_{count} = F\left(\frac{K_{top} \times B_{sz}}{n - n_{count}}\right) + b_{token}
\end{equation}

\textbf{Approximation and Solution.} To derive an analytical relationship for optimal resource allocation, we introduce two practical assumptions based on the workload characteristics of large-scale MoE models:

\begin{enumerate}
    \item \textbf{Communication Dominance:} The token dispatching group is significantly more resource-intensive than token count processing. Consequently, the majority of cores are allocated to communication ($n_{token} \gg n_{count}$), implying:
    \begin{equation}
        n - n_{count} \approx n
    \end{equation}
    
    \item \textbf{Negligible Constant Overhead:} Under high-load scenarios (large batch size and Top-$k$), the variable processing time dominates the constant overheads. Thus, we approximate $b_{count} \approx 0$ and $b_{token} \approx 0$.
\end{enumerate}

Assuming $F(x) = \alpha x$ and $G(x) = \beta x$ are linear functions, Eq. \ref{eq:balance} simplifies to:

\begin{equation}
    \beta \cdot \frac{N_{exp}}{n_{count}} \approx \alpha \cdot \frac{K_{top} \times B_{sz}}{n}
\end{equation}

Solving for $n_{count}$, we obtain:

\begin{equation}
    n_{count} \approx \frac{\beta}{\alpha} \cdot n \cdot \frac{N_{exp}}{K_{top} \times B_{sz}}
\end{equation}

The derivation leads to the following proportionality:
\begin{equation}
    n_{count} \propto \frac{N_{exp}}{K_{top} \times B_{sz}}
\end{equation}

This result indicates that to maintain optimal pipeline overlap, the number of cores allocated to token count processing ($n_{count}$) should be \textbf{directly proportional to the number of experts ($N_{exp}$)} and \textbf{inversely proportional to the token load ($K_{top} \times B_{sz}$)}. As the communication load increases, resources must be shifted from calculation to communication to prevent the dispatch group from becoming the bottleneck.

\section{Hardness Analysis}
\label{sec:hardness_proof}

\textbf{Theorem 1.} \textit{Program \eqref{eq:opt_model} is NP-Hard.}

\begin{proof}
We analyze the hardness by reducing from the \textbf{Subset Sum Problem (SSP)}. 
Consider a Subset Sum Problem Instance $\mathcal{I}_2 = \{a_0, a_1, \dots, a_{k-1}\}$, where $\sum_{i=0}^{k-1} a_i = 2B$.

We construct an instance $\mathcal{I}_n$ of the original scheduling problem consisting of a set of $k \cdot n$ tokens:
\begin{equation}
    \mathcal{I}_n = \left\{ 
    \begin{aligned}
        & (a_0, 0), \dots, (a_{k-1}, 0), \\
        & \underbrace{(B, 0), \dots, (B, 0)}_{(n-2) \times (B, 0)\text{'s}}, 
          \underbrace{(0, 0), \dots, (0, 0)}_{((k-1)(n-1)+1) \times (0, 0)\text{'s}} 
    \end{aligned}
    \right\}
\end{equation}

Consider the following decision problem: Does there exist an assignment $\{x_{ij}\}$ such that:
\begin{equation}
    \min_{\{x_{ij}\}} \max_{j} \left( \sum_{i \in M_j} \ell_i + \max_{i \in M_j} \{ \text{RTT}_i \} \right) \le B
\end{equation}

If, for $\mathcal{I}_n$, there exists an $n$-partition $M_1, M_2, \dots, M_n$ that satisfies the requirements of the original problem:
Because the total load is:
\begin{equation}
    (n-2) \cdot B + \sum_{i=0}^{k-1} a_i = (n-2) \cdot B + 2B = n \cdot B
\end{equation}
And due to the cardinality constraint:
\begin{equation}
    |M_1| = |M_2| = \dots = |M_n| = k
\end{equation}

Therefore, among $M_1, M_2, \dots, M_n$, there must be exactly $(n-2)$ partitions that take the form:
\begin{equation}
    \left\{ (B, 0), \underbrace{(0, 0), (0, 0), \dots, (0, 0)}_{(k-1) \times (0, 0)\text{'s}} \right\}
\end{equation}
Without loss of generality, let these be $M_3, M_4, \dots, M_n$ as shown above.

Consequently, $M_1$ and $M_2$ must strictly contain the terms from $\{(a_0, 0), \dots, (a_{k-1}, 0)\}$ and the remaining several $(0, 0)$ tokens.
Let $N_1 = \{ a_i \mid (a_i, 0) \in M_1 \}$ and $N_2 = \{ a_j \mid (a_j, 0) \in M_2 \}$.
We then have:
\begin{equation}
    \sum_{a_i \in N_1} a_i = B = \sum_{a_j \in N_2} a_j, \quad N_1 \cap N_2 = \emptyset
\end{equation}
This implies that $N_1, N_2$ constitute a valid solution to the requirements of the Subset Sum Problem instance $\mathcal{I}_2$.

In summary, to solve the Subset Sum Problem, one only needs to construct the corresponding original problem instance and solve it. Since the Subset Sum Problem is NP-Complete, the original decision problem is NP-Hard.
\end{proof}

\section{Supplementary Experiment}

\subsection{Extended Latency Breakdown}
\label{sec:appendix_latency_eval}

% latency (变化参数）
% bs = 32, 128; ranknum = 64, 256
\begin{figure}[t]
  \centering
  % 第一张子图
  \begin{subfigure}{0.88\columnwidth}
    \centering
    \includegraphics[width=0.88\linewidth]{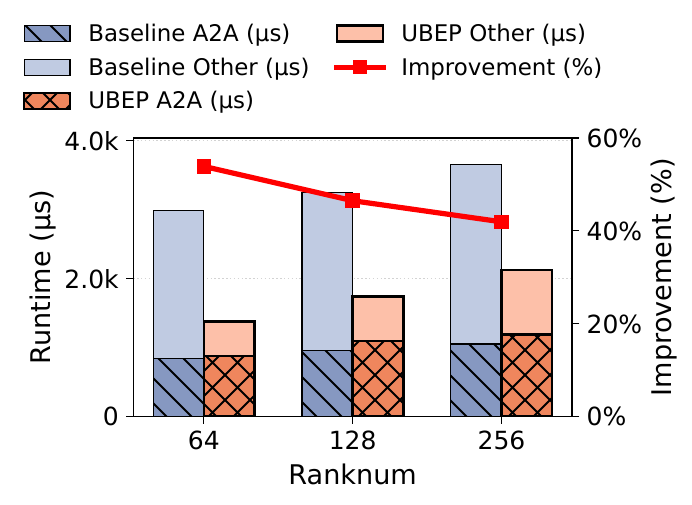}
    \vspace{-1em} 
    \caption{BS=32}
    \label{fig:appendix_latency_bs32}
  \end{subfigure}
  \begin{subfigure}{0.9\columnwidth}
    \centering
    \includegraphics[width=0.88\linewidth]{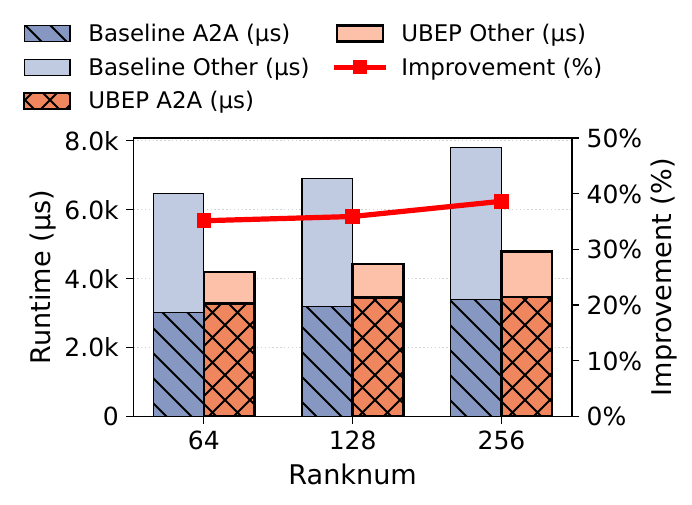}
    \vspace{-1em} 
    \caption{BS=128}
    \label{fig:appendix_latency_bs128}
  \end{subfigure}
  % Rank
  \begin{subfigure}{0.9\columnwidth}
    \centering
    \includegraphics[width=0.88\linewidth]{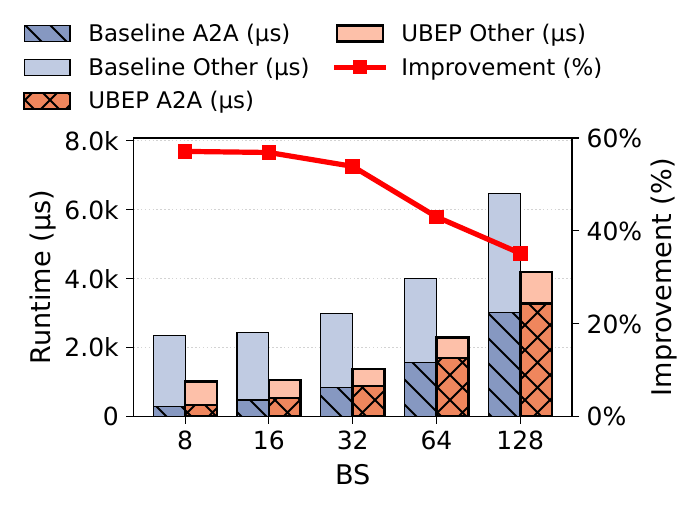}
    \vspace{-1em} 
    \caption{64 ranks}
    \label{fig:appendix_latency_ranknum64}
  \end{subfigure}
    \begin{subfigure}{0.9\columnwidth}
    \centering
    \includegraphics[width=0.88\linewidth]{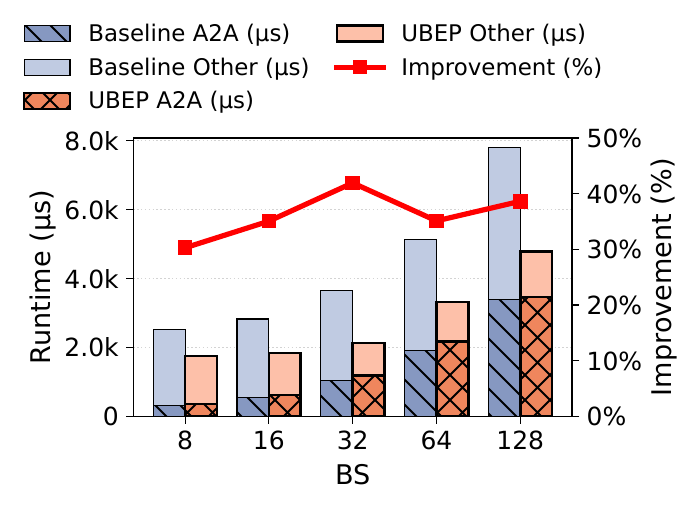}
    \vspace{-1em} 
    \caption{256 ranks}
    \label{fig:appendix_latency_ranknum256}
  \end{subfigure}
  \vspace{-1em} 
  \caption{Supplementary latency analysis of the MoE Dispatch primitive}
  \label{fig:appendix_latency}
  \vspace{-1em} 
\end{figure}

We extend the latency composition analysis in \S\ref{sec:ablation_study} to additional system configurations. These results examine whether \nickname exhibits consistent behavior across different batch sizes and cluster scales.

\noindent\textbf{Varying Batch Size.} Figure~\ref{fig:appendix_latency_bs32} and~\ref{fig:appendix_latency_bs128} show latency breakdowns for batch sizes of 32 and 128. \nickname reduces the dominant latency components by 35\%--54\% over the baseline for both smaller and larger batch sizes, confirming its efficiency across batch sizes.                                           
\noindent\textbf{Varying Cluster Scale.}                Figure~\ref{fig:appendix_latency_ranknum64} and~\ref{fig:appendix_latency_ranknum256} report results for 64 and 256 ranks.    
Despite increased All-to-All communication and synchronization overhead at larger scales, \nickname reduces latency by 30\%--42\% compared to the baseline, demonstrating robust scalability.

\subsection{Robustness to Routing Sparsity (Top-$k$)}
\label{sec:additional_experimental_parameters}

\begin{figure}[t]
    \centering
    \includegraphics[width=1.0\linewidth]{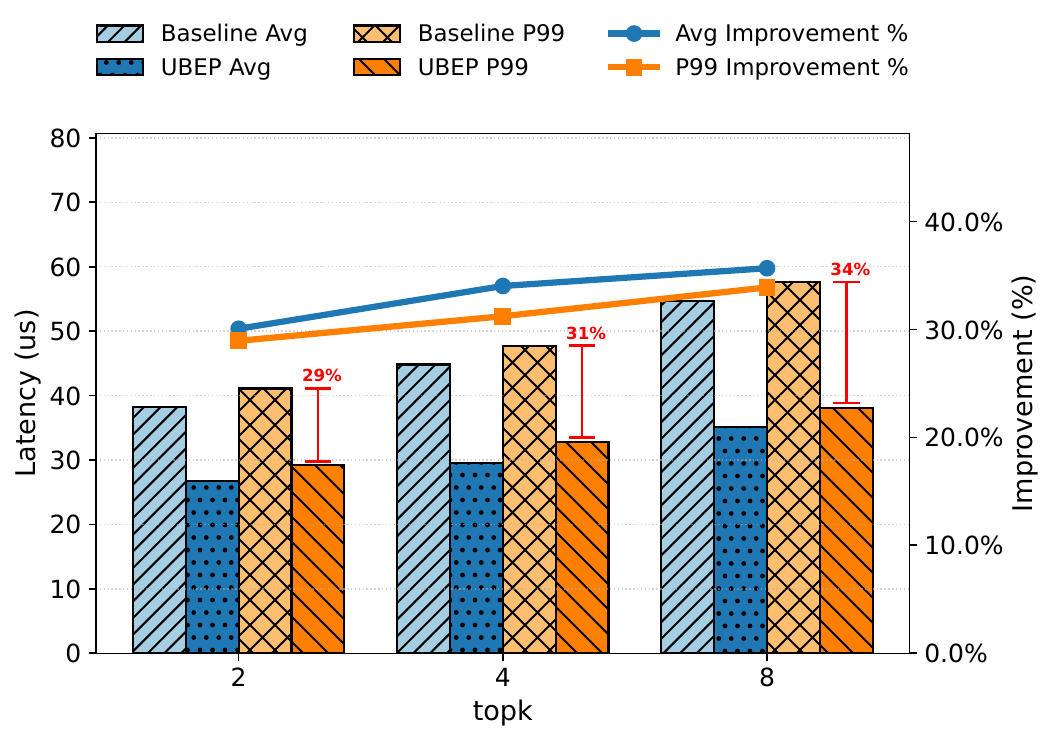}
    \vspace{-2em} 
    \caption{Performance scalability of different Top-$k$.}
    \label{fig:topk_performance}
    \vspace{-1em} 
\end{figure}

% \todo{title should change.}
To better test how \nickname's performance changes with parameters, we have added one more experiments in addition to \S\ref{sec:evaluation}. Figure~\ref{fig:topk_performance} presents the sensitivity analysis for the Top-$k$ parameter. Although increasing Top-$k$ results in a proportional multiplication of routing traffic, \nickname's performance improvement rate remains robust, maintaining a high range of 29\%--34\%. This stability under throughput pressure validates the robustness of our vectorized routing kernel in handling intensive memory accesses.

% % 分析 NMOE (模型复杂度) 
% \noindent\textbf{Model Scale.} Regarding the diversity of MoE architectures, Figure~\ref{fig:nmoe_performance} demonstrates that as the number of experts per token (NMOE) increases, the routing logic complexity grows exponentially. 

\subsection{Tuning Validation Resource Allocation}
\label{sec:hot_figure}

\begin{figure}[t]
  \centering
  
   \begin{subfigure}{1.0\columnwidth}
    \centering
    \includegraphics[width=1.0\linewidth]{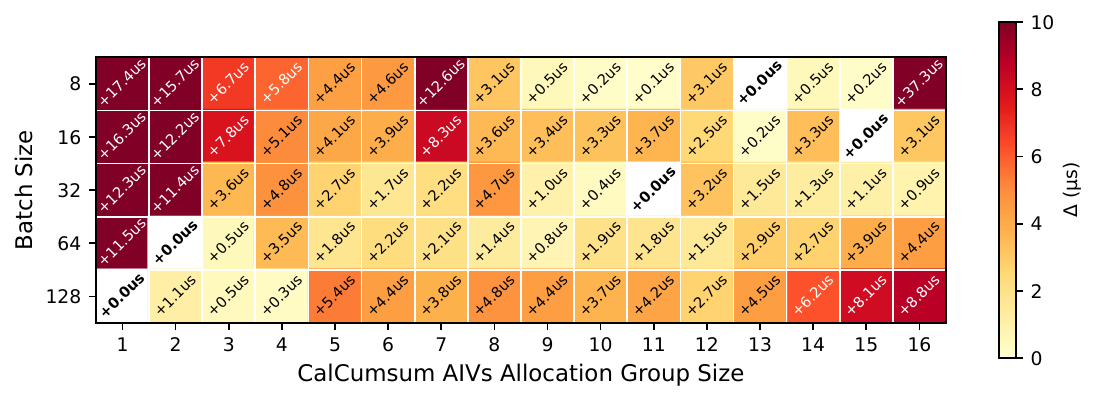}
    \caption{64 ranks}
    \label{fig:appendix_heatmap_rank64}
  \end{subfigure}
  
  \begin{subfigure}{1.0\columnwidth}
    \centering
    \includegraphics[width=1.0\linewidth]{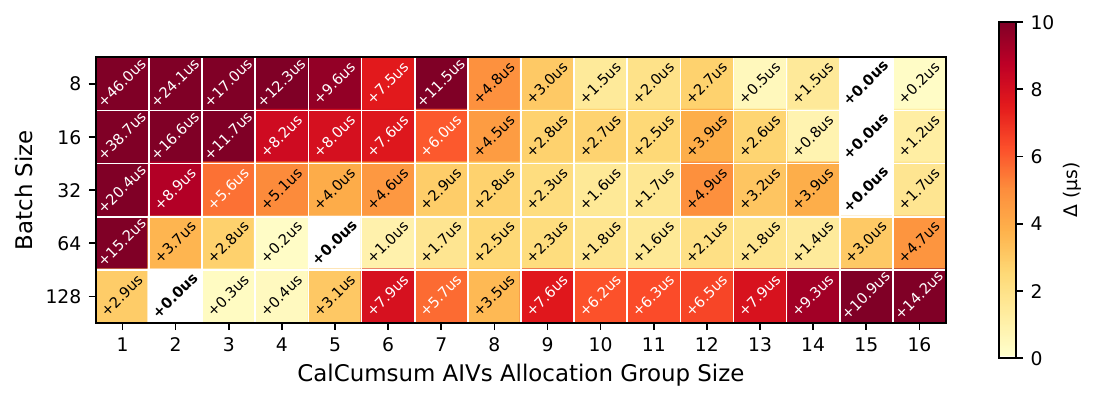}
    \caption{128 ranks}
    \label{fig:appendix_heatmap_rank128}
  \end{subfigure}
  % \vspace{0.5em} % 
  % \vspace{0.5em} % 

    \begin{subfigure}{1.0\columnwidth}
    \centering
    \includegraphics[width=1.0\linewidth]{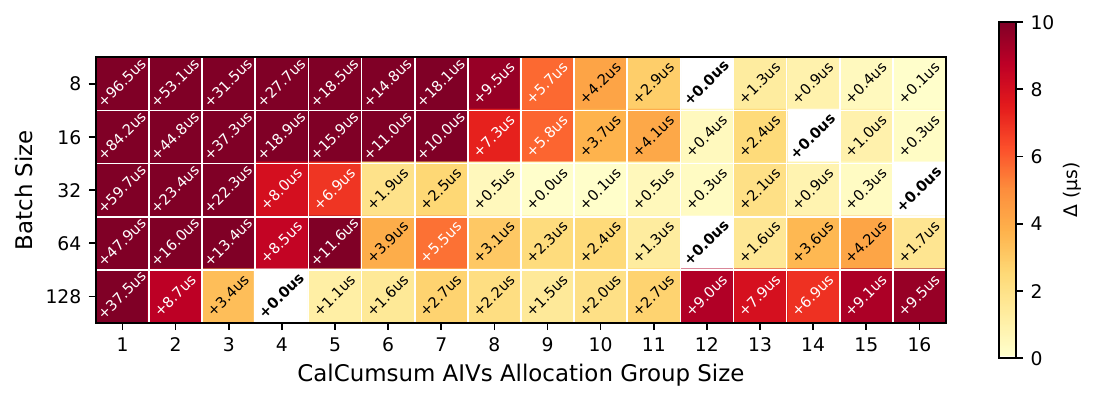}
    \caption{256 ranks}
    \label{fig:appendix_heatmap_rank256}
  \end{subfigure}
  \vspace{0em} % 调节两图之间的间距
   \caption{Heatmaps illustrating the relationship between the \texttt{CalCumSum} AIVs allocation group size, batch size, and varying rank numbers}
  \label{fig:appendix_heatmap}
  
\end{figure}

We sweep the \texttt{CalCumSum} AIV allocation group size across 64, 128, and 256 ranks to find the best setting for different combinations of rank count and batch size, as shown in Figure~\ref{fig:appendix_heatmap}. For each grid point, we normalize the latency of the best group size to 0 and report the extra latency of the other group sizes relative to the optimum; larger values indicate a larger deviation from the optimum.

Two trends are consistent. First, the best group size decreases as batch size grows. With larger batches, dispatch spends more time on communication, so \texttt{CalCumSum} verification is more likely to be overlapped by communication. Adding more \texttt{CalCumSum} AIVs then brings less benefit and can introduce extra scheduling and contention overhead, making fewer AIVs preferable. Second, the best group size increases with rank count. We fix one expert per NPU die, so more ranks mean more experts and more validation work. Allocating more \texttt{CalCumSum} cores increases parallelism and reduces stragglers on the verification path.
\end{sloppypar}
\end{document}